\def \S{{\bf S}}

\def\k{{\bf k}}
\def \Q{{\bf Q}}
\def\R{{\bf R}}

\def\shat{\hat{s}}
\def\that{\hat{t}}
\def\qhat{\hat{q}}

\def\gammahat{\hat{\gamma}}
\def\sbar{\bar{s}}

\def\Hhat{\hat{H}}

\def\qhat{\hat{q}}
\def\that{\hat{t}}
\def\sbar{\bar{s}}
\def\qbar{\bar{q}}
\def\Qbar{\bar{Q}}

\def\k{{\bf k}}
\documentclass[epj]{svjour}
\usepackage{graphicx,color,epsfig,rotate}
\usepackage{mathbbol}
\usepackage{amsfonts}
\begin{document}
\title{Quantum to classical transition in the ground state of a spin-$S$ quantum antiferromagnet}
\author{B. Danu\inst{1} \and  B. Kumar\inst{2}
}  
\offprints{Quantum to classical transition in the ground state of a spin-$S$} 
\institute{The Institute of Mathematical Sciences, C I T Campus,  Chennai 600113, India. \and School of Physical Sciences, Jawaharlal Nehru University, New Delhi 110067, India.}
\date{March 20, 2017}
\abstract{
We study a frustrated spin-$S$ staggered-dimer Heisenberg model on square lattice by using the bond-operator representation for quantum spins, and investigate the emergence of classical magnetic order from the quantum mechanical (staggered-dimer singlet) ground state for increasing $S$.  Using triplon analysis, we find the critical couplings for this quantum phase transition to scale as $1/S(S+1)$. We extend the triplon analysis to include the effect of quintet dimer-states, which proves to be essential for establishing the classical order (N\'eel or collinear in the present study) for large $S$, both in the purely Heisenberg case and also in the model with single-ion anisotropy. 
} 
\maketitle
\section{Introduction}
 The frustrated quantum antiferromagnets are a subject of great interest. The frustration arising due to competing interactions or lattice geometry, helped further by the quantum fluctuations, acts unfavourably towards classical magnetic order, and can give rise to exotic quantum paramagnetic 
states~\cite{Fmila2011}.  Many different materials are known to display frustrated magnetic behaviour in a variety of ways. For instance, CuGeO$_3$~\cite{Hase1993}, SrCu$_2$(BO$_3$)$_2$~\cite{Kageyama1999},CaV$_4$O$_9$~\cite{Taniguchi1995}, and  ZnCu$_3$(OH)$_6$Cl$_2$~\cite{Helton2007},  are some of the very well known materials of this kind. On the theoretical side, there is a lot of interest in exploring and understanding the exotic behaviour in variously motivated frustrated quantum spin models. 
 
The simplest way in which the frustrated antiferromagnets can exhibit quantum paramagnetism is through the formation of pairwise singlets, called dimers, in the ground state. The pioneering work of Majumdar and Ghosh showed this exactly in a spin-1/2 Heisenberg chain~\cite{MG}. Similarly, there is a strong numerical support for the spin-1/2 $J_1$-$J_2$ model on square lattice to have non-magnetic columnar dimer ground state in the range of $0.4< J_2/J_1<0.6$~\cite{Sachdev1990,Poilblanc1991,Oitmaa1996}. The dimer ground states have been shown to occur in many other frustrated models~\cite{SS,Klein,IndraniBose,BK2002,Bimla-BK2012,BK2005,Gelle-BK2008,Rakesh-BK2008,Rakesh-Dushyant-BK2009}. 
 
A dimer singlet ground state can also be realized by taking the antiferromagnetic exchange interaction between the spins of selected pairs to be stronger than the rest. For instance, in Fig.~\ref{fig:stg_model_fig} denoting a Heisenberg model on square lattice, the staggered pattern of thick nearest-neighbour bonds ($J^\prime$) is taken to be stronger compared to the other interactions ($J$ and $J_f$). The spin-1/2 case of this staggered-dimer model without $J_f$ has been studied by using the perturbative series expansion method~\cite{Singh1988},  exact diagonalization~\cite{Kruger2000}, coupled cluster method~\cite{Kruger2000}, and quantum Monte Carlo simulation~\cite{Wenzel2008,Jiang2012}. These studies found the staggered dimer ground state for strong enough $J^\prime/J$, which undergoes a transition to the N\'eel state as $J^\prime/J$ becomes weaker. A case for topological order in this model has also been made on the basis of tensor network studies~\cite{Huang2011}.  While the studies concerned with dimer states focus mostly on spin-1/2 problems, Darradi \emph{et al} made an interesting investigation of their dependence on spin quantum number, $S$, in the unfrustrated ($J_f=0$) staggered-dimer model~\cite{RDarradi2005}. Using exact diagonalization, coupled-cluster method and variational mean-field approach, they studied the $S$ dependence of the critical coupling for transition from staggered dimer to N\'eel phase in this model. Based on their calculations for upto $S=2$, they inferred that $(J/J^\prime)_c \propto 1/S(S+1)$.

The triplon analysis, based on the bond-operator representation of spins~\cite{Sachdev1990,Kumar2010}, is a nice analytical method for studying the ordering instabilities of the dimer singlet states. Using the spin-$S$ bond-operator representation of Kumar~\cite{Kumar2010}, in this paper, we do mean-field triplon analysis of the frustrated staggered-dimer model on square lattice (see Fig.~\ref{fig:stg_model_fig}). We find the ground state of this Heisenberg model to undergo continuous phase transition from the staggered dimer to N\'eel or collinear state, and the critical couplings for both the transitions to scale precisely as $1/S(S+1)$. This lends support to the earlier finding for $J_f=0$~\cite{RDarradi2005}. Interestingly, for strong enough frustration (when $J_f \sim J/2$, the Majumdar-Ghosh value) 
we also find the staggered dimer phase to survive in the large $S$ (classical) limit. It looks a bit unusual, and turns out to be an artefact of considering only the singlet and triplet dimer states in the minimal low-energy description. We improve the triplon analysis by including the effect of quintet states, which gives the expected classical magnetic order for large spins. 
We also get the classical behaviour in the large $S$ limit by adding an infinitesimally small axial single-ion anisotropy (that is ever-present \footnote{In high spin magnetic materials, it is pretty common to have axial anisotropies, as has been reported in many materials such as DTN [NiCl$_2$4SC(NH$_2$)$_2$]~\cite{Zvyagin2007}, NENC [Ni(C$_2$H$_8$N$_2$)$_2$Ni(CN$_4$)]~\cite{Orend1995}, CsNiCl$_3$~\cite{Buyers1985}, LiFePO$_4$~\cite{JiyingLi2006} {\it etc}.}).
Through this study of the $S$ dependence of quantum phase transitions in the staggered-dimer model, we present an interesting perspective on the approach to classical behaviour from a quantum state with increasing $S$, and extend the triplon analysis in a simple way to be applicable to the higher spin dimerised Heisenberg antiferromagnets.

This paper is organized as follows. In Sec.~\ref{A quantum spin-$S$ model}, we state the staggered-dimer model on square lattice, and carry out the simplest mean-field triplon analysis of its purely Heisenberg case, using bond-operator representation in the reduced space of singlet and triplet dimer states. Then, in Sec.~\ref{stq}, we bring quintet states into the discussion. In particular, in Sec.~\ref{stq-mft}, we extend the triplon analysis to include the effect of quintets in a mean-field approach that is nearly as simple as the minimal triplon analysis, but makes significant physical improvement to the results of the calculation. We also point out that how it is further extendable to include the higher dimer states such as septets, octets and so on. In Sec.~\ref{Model with single-ion anisotropy}, we study the effect of single-ion anisotropy with increasing $S$. We then conclude with a summary in Sec.~\ref{Summary:}.  

\section{Staggered-dimer model on square lattice} \label{A quantum spin-$S$ model} 
\begin{figure}[ht]
\begin{center}
\includegraphics[width=0.98\columnwidth]{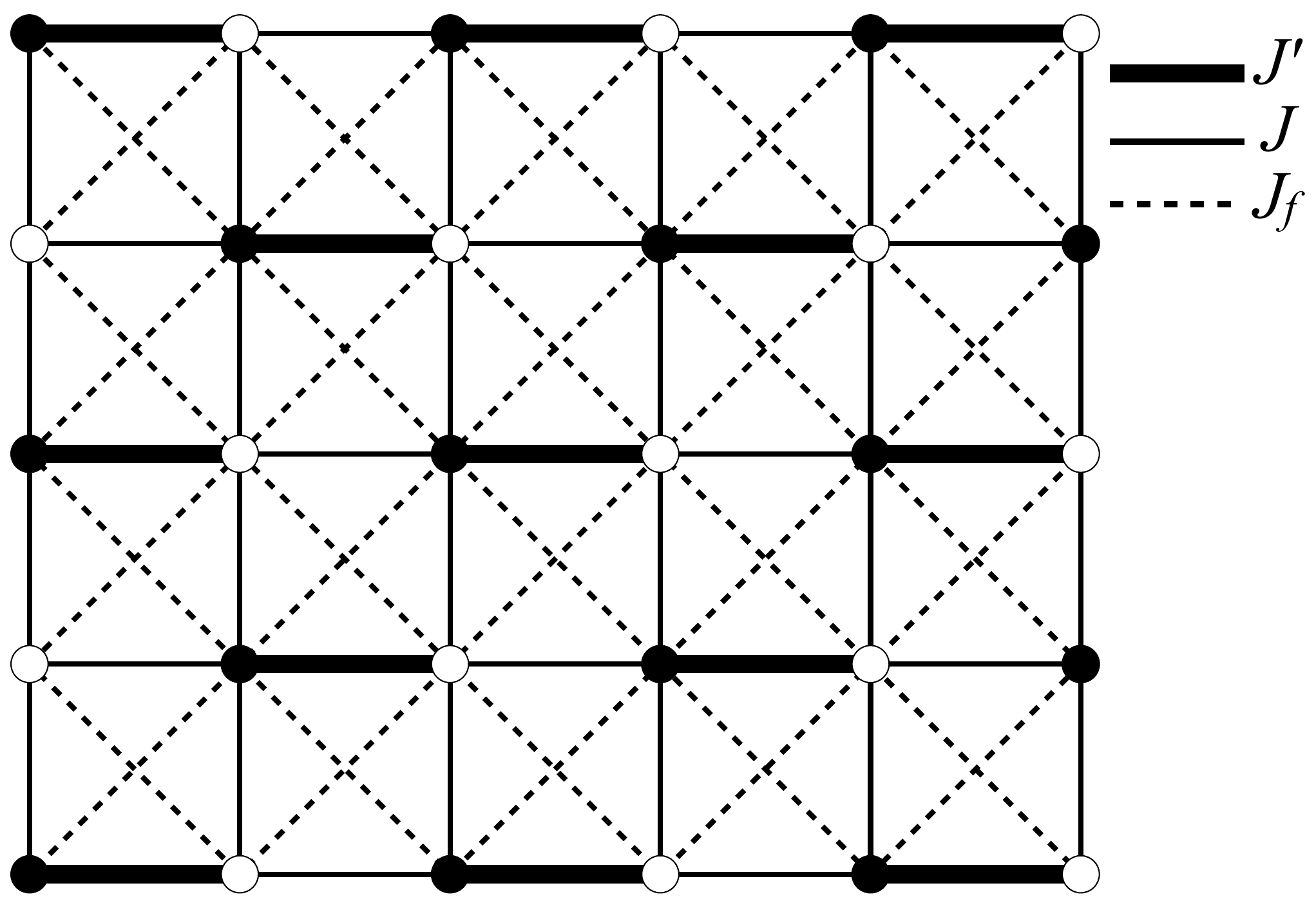}
 \caption {The staggered-dimer model on square lattice. 
 }
 \label{fig:stg_model_fig}
\end{center}
\end{figure}

The Hamiltonian of the spin-$S$ staggered-dimer model on square lattice depicted in Fig.~\ref{fig:stg_model_fig} can be written as follows.
\begin{eqnarray}\label{stg_mdl}
\Hhat&=&\sum _{{\bf R}}\Big[ J^{\prime} {\bf S}_{{\bf R},1}. {\bf S}_{{\bf R},2}-D({\bf S}^{2}_{{\bf R},1z}+{\bf S}^{2}_{{\bf R},2z})\nonumber\\
&&+J ({\bf S}_{{\bf R},2}.{ \bf S}_{{\bf R}+2 a\hat x,1}+{\bf S} _{{\bf R},2}.{\bf S}_{{\bf R}+a \hat {x} \pm a \hat{y},1})\nonumber\\
&&+J_f({ \bf S}_{{\bf R},1}.{\bf S}_{{\bf R}+a\hat {x} \pm a\hat{y},1}+{\bf S}_{{\bf R},2}.{\bf S}_{{\bf R}+a\hat{x}\pm a\hat{y},2})\Big]
\end{eqnarray}
Here, $\bf{R}$ is summed over the positions of the dimers, and the labels $1$ and $2$ denote the spins of a dimer.  The $J^{\prime}$ is the intra-dimer exchange interaction. The $J$ and $J_f$ are the interactions between the spins of the nearest dimers.  All of these exchange interactions are antiferromagnetic. The single-ion anisotropy, $D$, is taken to be a positive number.  

The purely Heisenberg case ($D=0$) of this model would clearly form the spin-gapped staggered-dimer singlet ground state for $J^{\prime}\gg J, J_f$. Here, the dimerization is not spontaneous, but induced by the stronger $J^{\prime}$. Besides a strong $J^\prime$, the frustration due to competing $J$ and $J_{f}$ also facilitates the non-magnetic dimer ground state. However, the N\'eel (or collinear) antiferromagnetic (AFM) order would be realized in the ground state for sufficiently strong $J$ (or $J_f$). The anisotropy, $D$, also tends to support magnetic order. Clearly, here there is an ample scope for competition between the quantum disordered staggered-dimer phase and the classical N\'eel or collinear ordered phases. We study this competition as a function of the spin quantum number, $S$, by doing the stability analysis of the staggered-dimer phase using bond operators. The Heisenberg case of $\Hhat$ with $D=0$ is discussed in the following subsections and in Sec.~\ref{stq}, and the model for $D\neq 0$ is studied thereafter.  
\subsection{Bond-operator representation  of spin-$S$ operators}
The systems where the pairs of spins (dimers) act as basic units, the bond-operator representation of spins has been found to be very useful in doing calculations. This approach was pioneered by Sachdev and Bhatt for studying quantum phase transitions in dimerized spin-1/2 quantum antiferromagnets~\cite{Sachdev1990}. For the case of general spin-$S$, the bond operator representation was derived by Kumar~\cite{Kumar2010}. The basic idea is to construct the spin eigenstates of a dimer, and associate with each one of them a distinct bosonic creation operator called bond operator. For a pair of spin-$S$ interacting via exchange interaction, $\S_1\cdot\S_2$, the suitable dimer states are the total-spin eigenstates. That is, the spin singlet: $|s\rangle$, triplet: $|t_{m_1}\rangle$, quintet: $|q_{m_2}\rangle$, and so on. The corresponding bond operators are defined as:
\begin{eqnarray*}
\big|s \rangle &=&\shat^{\dagger}|0\rangle,\quad
 \big|t_{m_{1}}\rangle =\that^{\dagger}_{m_{1}}\big|0\rangle,\quad
 \big|q_{m_{2}}\rangle=\qhat^{\dagger}_{m_{2}}\big|0\rangle. 
\end{eqnarray*}
Here, $\big|0 \rangle$ denotes the vacuum of the bosonic Fock space. In order to describe the finite-dimensional spin Hilbert space of the dimer, these bond operators in the Fock space are required to satisfy the constraint: $\shat^{\dagger}\shat+\that^{\dagger}_{m_{1}}\that_{m_{1}}+\qhat^{\dagger}_{m_{2}}\qhat_{m_{2}}+\cdots=1$, where `$\cdots$' is meant to denote the contributions from all the higher total-spin dimer states upto $2S$. The exchange operator on the dimer can now be written as:
\begin{eqnarray}
 \S_{1}.\S_{2}&=& -S(S+1)\left[\shat^{\dagger}\shat+\that^{\dagger}_{m_{1}}\that_{m_{1}}+\qhat^{\dagger}_{m_{2}}\qhat_{m_{2}}+\cdots \right] \nonumber \\
 && +\left[\hat{t}^{\dagger}_{m_{1}}\hat{t}_{m_{1}}+3 \qhat^{\dagger}_{m_{2}}\qhat_{m_{2}}+\cdots \right].
\end{eqnarray}

To write the interactions between the spins of different dimers, one needs the bond-operator representation of the individual spins. At the very least, it can constructed in the restricted space of singlet and triplet, and can also be extended further to include quintet or higher states, if necessary.
 The bond operator representation of $\S_{1}$ and $ \S_{2}$ in the restricted space of singlet and triplet has the following form~\cite{Kumar2010}. 
  \begin{eqnarray}
 \S_{(1,2)\alpha}&\approx&\pm\sqrt{\frac{S(S+1)}{3}}(\shat^{\dagger} \that_{\alpha}+\that^{\dagger}_{\alpha}\shat)-\frac{i}{2}  \epsilon_{\alpha \beta \gamma} \that^{\dagger}_{\beta}\that_{\gamma},   \label{eq:rep-st1}
 \end{eqnarray}
Here, $\alpha=x,y,z$, and likewise for $\beta$ and $\gamma$. The $ \epsilon_{\alpha \beta \gamma}$ is the totally antisymmetric tensor, and the signs $+$ and $-$ correspond to the spin-label 1 and 2, respectively. 
\subsection{Mean-field triplon analysis of the Heisenberg case}
  \label{sub:Triplon mean-field Hamiltonian:}
The staggered-dimer {\em Heisenberg} model, $\Hhat$ of Eq.~(\ref{stg_mdl}) with $D=0$, in the bond-operator representation turns into a model of interacting triplets and singlets, which is hard to make progress with. To simplify it, we make the following standard approximations~\cite{Sachdev1990,Rakesh-BK2008}. We treat the singlet bond-operators (for the staggered singlet bonds of Fig.~\ref{fig:stg_model_fig}) in the mean-field approximation as: $\langle \hat s\rangle$$=$$\langle \hat s^{\dagger}\rangle$$=$$\bar s$. It now becomes a model of interacting triplet excitations with a mean singlet background. To keep it simple, we also neglect the interaction between triplets, and impose the local constraint on the bond-operators through an average Lagrange multiplier, $\lambda$. The resulting Hamiltonian is bilinear in triplet bond-operators, called {\em triplons}. Finally, by doing the Fourier transformation: $\that_{\bf{R}\alpha} = \frac{1}{\sqrt{N_d}}\sum_{\bf{k}}e^{i \k.\R}\that_{\bf{k}\alpha}$, we get the following effective triplon dynamics.
\begin{eqnarray} \label{eq:hmf_kspc}
\Hhat_{t}&= &N_d\left[J^{\prime}-J^{\prime}S(S+1)-\frac{5}{2}\lambda +\bar{s}^2 (\lambda-J')\right]  \nonumber\\ &&+\frac{1}{2}\sum_{\bf{k},\alpha}\bigg\{\Big[\lambda-\bar{s}^2 S(S+1) \xi_{\bf{k}} \Big]\left(\hat t^\dag_{\bf{k}\alpha} \hat t_{\bf{k} \alpha} + \hat t_{-\bf{k}\alpha}\hat t^\dag_{-\bf{k}\alpha}\right) \nonumber \\
&&- \bar{s}^2 S(S+1)\xi_{\bf{k}} \left(\hat t^\dag_{\bf{k}\alpha} \hat t^\dag_{-\bf{k}\alpha} + \hat t_{-\bf{k}\alpha}\hat t_{\bf{k}\alpha}\right) \bigg\} .
\end{eqnarray}
Here, $\bf{k}$ lies in the Brillouin zone  of the staggerd-dimer square lattice, $N_{d}$ is the total number of dimers, and $\xi_{\bf k}=\frac{2J}{3}\cos (2k_{x}a)+\frac{4J}{3}\cos(k_{x}a) \cos(k_{y}a)-\frac{8J_f}{3}\cos(k_{x}a)\cos(k_{y}a)$.

The triplon Hamiltonian, $\Hhat_t$, can be diagonalized by the Bogoliubov transformation for bosons, $ \hat{t}_{\bf{k}\alpha}$=$\hat\gamma_{\bf{k}\alpha}\cosh {\theta_{\bf{k}}}\\-\hat\gamma_{-\bf{k}\alpha}^{\dagger}\sinh \theta_{\bf{k}}$, with $\theta_{\bf k} = \frac{1}{2}\tanh^{-1}\{-\bar{s}^2 S(S+1)\xi_{\bf{k}}/[\lambda-\bar{s}^2 S(S+1)\xi_{\bf{k}}]\}$. The $\Hhat_t$ in terms of the quasiparticle operators, $\hat\gamma_{\bf{k}\alpha}$, has the following diagonal form.
\begin{eqnarray}
\Hhat_{t}& =& e_{0} N_{d} + \sum_{\bf{k},\alpha}E_{\bf{k}}\left(\hat\gamma^\dag_{\bf{k}\alpha}\hat\gamma_{\bf{k}\alpha} +\frac{1}{2}\right) 
\end{eqnarray}
Here, $e_{0}=J^\prime-J^\prime S(S+1)-\frac{5}{2}\lambda+\bar{s}^{2}(\lambda-J^\prime)$, and  $E_{\bf{k}}=\sqrt{\lambda[\lambda-2\bar{s}^{2}S(S+1)\xi_{\bf{k}}]}\ge0$ is the triplon quasiparticle dispersion. The ground state energy per dimer is given by
  \begin{eqnarray}
 e_{g} \big[\lambda,\bar{s}^{2}\big]&=&e_{0}+\frac{3}{2N_{d}}\sum_{\bf{k}} E_{\bf{k}}.
 \end{eqnarray}
Minimization of the ground state energy density, $e_{g}$, with respect to the mean-field parameters, ${\lambda}$ and $\bar{s}^{2}$, gives the following equations.
 \begin{eqnarray}
  \bar{s}^{2}&=&\frac{5}{2}-\frac{3}{2 N_{d}}\sum_{\bf{k}}\frac{\lambda-\bar{s}^{2}S(S+1)\xi_{\bf{k}}}{E_{\bf{k}}} \label{sbar}\\   
  \lambda&=&J^{\prime}+ \frac{3\lambda S(S+1)}{2N_d}\sum_{\bf{k}}\frac{\xi_{\bf{k}}}{E_{\bf{k}}} \label{lambda}
 \end{eqnarray}
We determine $\bar{s}^{2}$ and  $\lambda$ 
by self-consistently solving the Eqs.~(\ref{sbar}) and~(\ref{lambda}) in the space of $J/J^\prime$ and $J_f/J^\prime$. In the staggered-dimer phase, the quasiparticle gap is non-zero. However, for some critical values of $J/J^\prime$ and $J_f/J^\prime$, this gap would vanish. The contours of such critical points define the boundaries of the staggered-dimer phase. On the other side of this phase boundary, the condensation of triplons at some wavevector $\bf{Q}$, given by $E_{\bf Q}=0$, determines the antiferromagnetic order. The vanishing of the gap also fixes the value of $\lambda$ at $\lambda^{\ast}=2  \bar{s}^{2} S(S+1)\xi_{{\bf Q}}$. 
\begin{figure*}[ht]
\begin{center}
\includegraphics[height=2.35in]{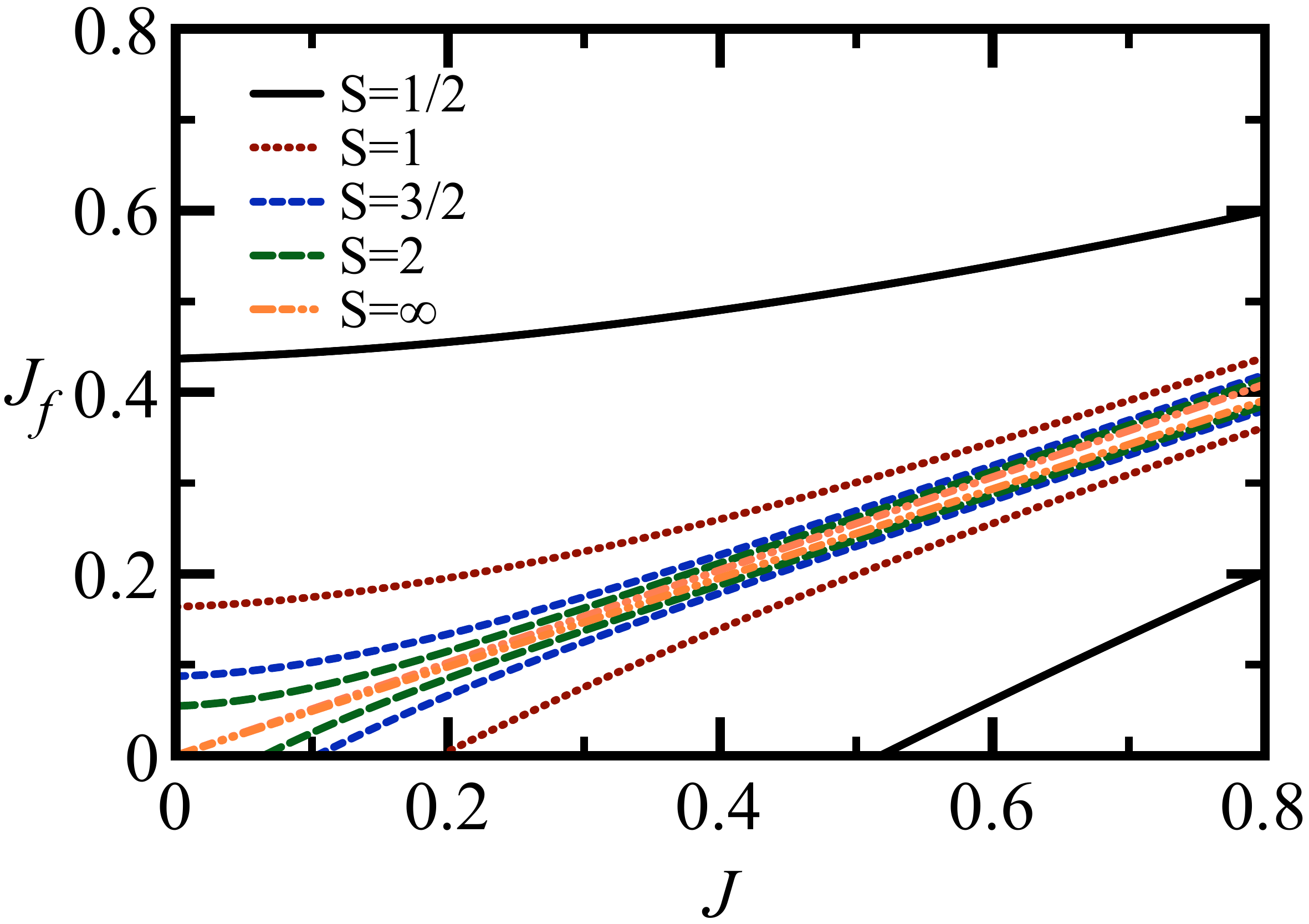}
\includegraphics[height=2.35in]{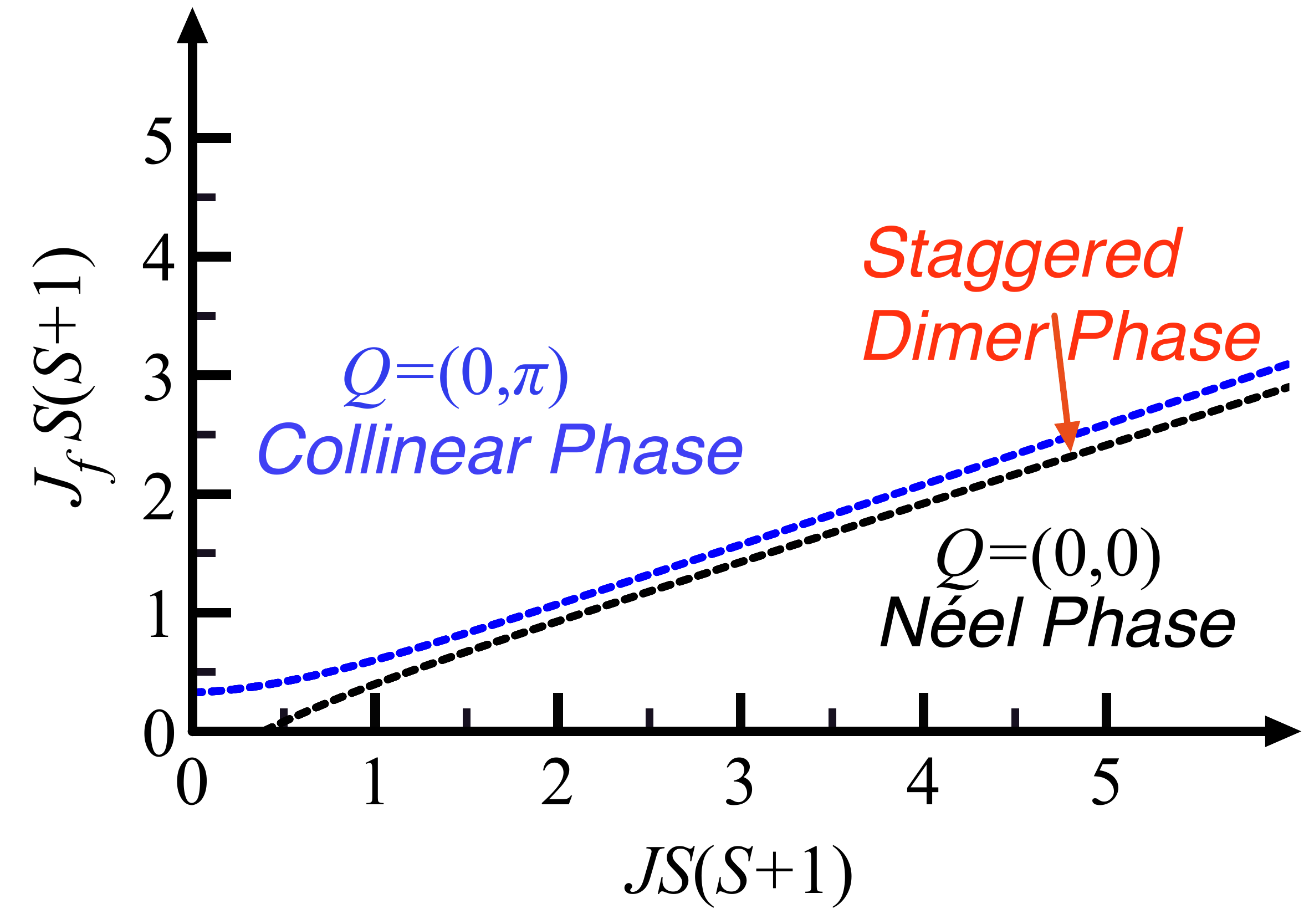}
\caption {Left: The mean-field  quantum phase diagram in the $J$-$J_f$ plane (for $J^{\prime}=1$ and $D=0$). The region above the upper phase boundaries is the collinear phase, below the lower phase boundary is the N\'eel phase, and between the two is the staggered-dimer phase. For $S$$=$$ \infty$,  the phase boundaries are $J_{f}$$=$$0.4890J$  and $J_{f}$$=$0.5109$J$ for N\'eel and collinear transition, respectively. Right: The quantum phase diagram in the space of rescaled parameters.}
 \label{fig:phs_s}
\end{center}
\end{figure*}
 \subsubsection{ Quantum phase diagram}
 \label{sub:Calculations and results:}
We set the intra-dimer interaction $J^\prime$ to 1, and vary $J$ and $J_f$ as free parameters. In the present case, $D=0$. We now describe the quantum phase diagram of the staggered-dimer Heisenberg model on square lattice in the $J$-$J_f$ plane, as obtained from the mean-field triplon analysis.

 From our calculations, we find two different ordering wavevectors at which the triplon gap vanishes. It is ${\bf Q}_{N}=(0,0)$  that gives N\'eel AFM order for sufficiently strong $J$, and  ${\bf Q}_{C}=(0,\pi)$ that gives collinear AFM order for strong enough $J_f$. We determine the phase-boundaries between the staggered-dimer and the two ordered AFM phases from the following equation. 
 \begin{eqnarray}
 \xi_{{\bf Q}}\left[5- \frac{3}{N_d}\sum_{{\bf k}}\sqrt{\frac{\xi_{{\bf Q}}}{(\xi_{{\bf Q}}-\xi_{{\bf k}})}}\right]&=&\frac{3}{2S(S+1)}.\label{sifnty}
\end{eqnarray} 
The quantum phase diagram, thus obtained, is presented in Fig.~\ref{fig:phs_s}. In Eq.~({\ref{sifnty}), the interaction parameters, $J$ and $J_f$, and the spin quantum number, $S$, appear separately on the left and right hand side, respectively. It clearly implies that the phase boundaries for all the different values of $S$ (in the left panel of Fig.~\ref{fig:phs_s}) will collapse onto  two lines only (one for dimer to N\'eel and the other for dimer to collinear transition) in the space of rescaled interactions, $JS(S+1)$ and $J_fS(S+1)$, as shown in the right panel of Fig.~\ref{fig:phs_s}. 

For $J_{f}=0$, the quantum critical point for staggered-dimer to N\'eel transition is given by $J_{c}\approx0.388/S(S+1)$. For $J=0$, the critical $J_f$ for transition to the collinear phase is $J_{fc}\approx0.3277/S(S+1)$. For $J_f=0$, we can compare the results of our calculations with those of others obtained by different methods. For $S=1/2$, we get $J_{c}=0.518$, while it is around $0.4$ from other numerical calculations~\cite{Singh1988,Wenzel2008,Jiang2012}. Take, for instance, the values of Darradi \emph{et al}~\cite{RDarradi2005} for $J_{c}$ $\approx$$(0.4545, 0.1818, 0.0990)$ obtained from  exact diagonalization method, and compare with $J^{c} \approx$ (0.5188, 0.194, 0.1038) from our calculation for $S=1/2, 1$ and $3/2$, respectively. The agreement is quite reasonable, and seems to get better as $S$ increases. Moreover, in the limit $S\rightarrow\infty$, when either $J$ or $J_{f}$ is zero (unfrustrated cases), the ordered classical phases are the only possibility. However, when both $J$ and $J_f$ are non-zero, the large $S$ limit turns out to be surprising, as discussed below.

In the limit $S\rightarrow \infty$, the right hand side of Eq.~(\ref{sifnty})  vanishes, which then gives two limiting phase boundaries: $J_{f}=0.4891J$ for staggered-dimer to N\'eel transition, and $J_{f} = 0.5109J$ for the transition to collinear phase. A small difference in their slopes leaves a tiny region of staggered-dimer phase between the two limiting phase boundaries (see Fig.~\ref{fig:phs_s}). It presents a puzzling situation for a Heisenberg problem in which a quantum mechanical (staggered-dimer) state, helped by strong frustration, survives into the classical limit. Quite possibly, this situation is an artefact of the oversimplified low-energy triplon dynamics, and can be resolved by including the corrections due to higher spin dimer-states. Below we particularly study the effect of quintets on the effective triplon dynamics.
{\color{black} \section{Quintet corrections to triplon analysis}\label{stq}
To include quintets in the effective low-energy dynamics of the spin model, we consider the bond-operator representation in the subspace of singlet, triplet and quintet dimer-states, as given in Ref.~\cite{Kumar2010}.
\begin{eqnarray}
\label{latticeS1}
S^z_{1,2} &\approx& \pm\sqrt{\frac{{\tt N}_t}{2{\tt N}_s}} \left( \shat^\dag\that^{ }_0 + \that^\dag_0 \shat \right) \pm\sqrt{\frac{{\tt N}_q}{{\tt N}_t}}\bigg[\frac{1}{\sqrt{3}}\left(\that^\dag_0\qhat^{ }_0 + \qhat^\dag_0\that^{ }_0\right) \nonumber \\
& +& \frac{1}{2}\left(\that^\dag_1\qhat^{ }_1 + \qhat^\dag_1\that^{ }_1+ \that^\dag_{\bar{1}} \qhat^{ }_{\bar{1}} + \qhat^\dag_{\bar{1}}\that^{ }_{\bar{1}}\right)\bigg] + \frac{1}{2}\Big(\that^\dag_1\that^{ }_1 - \that^\dag_{\bar{1}}\that^{ }_{\bar{1}} \nonumber \\
& +& \qhat^\dag_1\qhat^{ }_1 - \qhat^\dag_{\bar{1}}\qhat^{ }_{\bar{1}}\Big) + \left(\qhat^\dag_2\qhat^{ }_2 - \qhat^\dag_{\bar{2}}\qhat^{ }_{\bar{2}}\right) 
\end{eqnarray}
\begin{eqnarray}
\label{latticeS2}
S^+_{1,2} & \approx & \pm\sqrt{\frac{{\tt N}_t}{{\tt N}_s}} \left( \shat^\dag\that^{ }_{\bar{1}} - \that^\dag_1 \shat \right) \pm\sqrt{\frac{{\tt N}_q}{{\tt N}_t}}\bigg[ \left(\that^\dag_{\bar{1}}\qhat^{ }_{\bar{2}} - \qhat^\dag_2\that^{ }_1\right) \nonumber \\
& & + \frac{1}{\sqrt{2}}\left(\that^\dag_0\qhat^{ }_{\bar{1}} - \qhat^\dag_1\that^{ }_0\right)+\frac{1}{\sqrt{6}}\left(\that^\dag_1 \qhat^{ }_0 - \qhat^\dag_0\that^{ }_{\bar{1}}\right)\bigg]\nonumber \\
& & + \frac{1}{\sqrt{2}}\left(\that^\dag_1\that^{ }_0 + \that^\dag_0\that^{ }_{\bar{1}}\right) + \sqrt{\frac{3}{2}}\left(\qhat^\dag_1\qhat^{ }_0 + \qhat^\dag_0\qhat^{ }_{\bar{1}}\right) \nonumber  \\
&& + \left(\qhat^\dag_2\qhat^{ }_1 + \qhat^\dag_{\bar{1}}\qhat^{ }_{\bar{2}}\right)
\end{eqnarray}
where ${\tt N}_s=(2S+1)$, ${\tt N}_t=2S(S+1)(2S+1)/3$, and ${\tt N}_q=2S(S+1)(2S-1)(2S+1)(2S+3)/15$. 

Using Eqs.~(\ref{latticeS1}) and~(\ref{latticeS2}), our Heisenberg model would look like $\Hhat=\Hhat_{t}+\bar{s} \Hhat_{\hat{t}\hat{t}\hat{q}}(S^2)+\Hhat_{\hat{t}\hat{t}\hat{q}\hat{q}}(S^2)+\Hhat_{\hat{t}\hat{t}\hat{t}\hat{t}}+\Hhat_{\hat{q}\hat{q}\hat{q}\hat{q}}$, of which the first term is the triplon model worked out in the previous section. The next significant contribution (in powers of $S$) comes from $\Hhat_{\hat{t}\hat{t}\hat{q}}$ and $\Hhat_{\hat {t}\hat{t}\hat{q}\hat{q}}$. Below we discuss the corrections to triplon dynamics from these terms involving quintet operators. First we compute the corrections using second order perturbation theory, which is however limited in its scope. More importantly, in Sec.~\ref{stq-mft}, we extend the triplon analysis to include the quintets in a simple way, which nicely resolves the issue with the classical limit within triplon mean-field theory, and is also extendible in principle to include the higher spin dimer-states.}
\subsection{\color{black} Second order perturbation theory in gapped phase} \label{energy correction}
The simultaneous coupling of the triplets to the singlet background and the quintet excitations will allow $\bar{s}{\hat H}_{\hat{t}\hat{t}\hat{q}}$ to contribute more directly to the low-energy dynamics of the dimer phase, as compared to the other terms involving quintets. Therefore, here we calculate the second order perturbative correction only due to $\bar{s}{\hat H}_{\hat{t}\hat{t}\hat{q}}$ to the ground state energy, and compute the change in the quantum critical points. We can write this term on a lattice as:
\begin{eqnarray}
\Hhat_{\bar{s}\hat{t}\hat{t}\hat{q}}=-\frac{\bar{s}}{4}\sqrt{\frac{\tt N_q}{\tt N_s}}\sum_{m=-2,\cdots,2}\sum_{i,\delta}\left[q^{\dagger}_{i,n} \hat{T}^{[m]}_{i,i+\delta} +h.c.\right]
\end{eqnarray}
where $\hat{T}_{i,i+\delta}=\that_{i,\alpha}(\that_{i+\delta,\beta}+ \that^\dagger_{i+\delta,\beta})$~\cite{Ganesh2011}. The second order correction to ground state energy can be written as: 
\begin{eqnarray*}
\Delta E^S_{\bar{s}\hat{t} \hat{t}\hat{q}} &=&\frac{ \bar{s}^2}{16}\frac{\tt N_q}{\tt N_s}\sum_{m,m^\prime}\sum_{h\ne0} \times \\
&& \sum_{i,\delta,i^\prime,\delta^\prime}\frac{\langle 0|\hat q_{i^\prime,m^\prime}\hat {T}^{[m]\dagger}_{_{i^\prime,i^\prime+\delta^\prime}} |h\rangle\langle h|\hat q^\dagger_{i,m} \hat {T}^{[m]}_{i,i+\delta}|0\rangle}{E_0-E_{h}} \nonumber
\end{eqnarray*}
 where the summation over $\sum_{h\ne0}$ represents the sum over all the excited states $|h_{m_j}\rangle$.
   In the ${\bf k}$-space the composite triplet operator $\sum_{\delta}\hat{T}^{[m]}_{i,i+\delta}$ can be written as, $\sum_{\delta}\hat{T}^{[m]}_{i,i+\delta}=\frac{1}{N_d}\sum_{{\bf p},{\bf k}}\hat{T}^{[m]}_{-{\bf k}+{\bf p},{\bf k}}e^{i{\bf p}.r}\epsilon_{\bf k}$.  Treating the quintet excitations as local with the constraint ($i^\prime,m^\prime=i,m$), we get 
  \begin{eqnarray}
 \label{pertubationEk}
 {\Delta} E^S_{\bar{s}\hat{t} \hat{t}\hat{q}}=\frac{1}{N_d}\frac{\tt N_q}{16 \tt N_s}\bar s^2\sum_{m}\sum_{n\ne0}\sum_{\bf p, k}\frac{|\langle n|\hat {T}^{[m]}_{-{\bf k}+{\bf p},{\bf k}}\epsilon_{\bf k}|0\rangle|^2}{E_0-E_n}.
\end{eqnarray}
The operator  $\hat {T}^{[m]}_{-{\bf k}+{\bf p}}$ for $m=-2,\cdots, 2$ is defined in the following way. 
\begin{eqnarray*}
\hat {T}^{[\pm 2]}_{-{\bf k}+{\bf p,\bf k}}&=&\hat{t}_{(-{\bf k}+{\bf p})_x}(\hat {t}_{{\bf k}_x}+\hat{t}^{\dagger}_{-{\bf k}_x})-\hat{t}_{(-{\bf k}+{\bf p})_y}(\hat {t}_{{\bf k}_y}+\hat{t}^{\dagger}_{-{\bf k}_y})\\&&\mp i\hat{t}_{(-{\bf k}+{\bf p})_x}(\hat {t}_{{\bf k}_y}+\hat{t}^{\dagger}_{-{\bf k}_y})\mp i\hat{t}_{(-{\bf k}+{\bf p})_y}(\hat {t}_{{\bf k}_x}+\hat{t}^{\dagger}_{-{\bf k}_x})\nonumber \\
\hat {T}^{[\pm 1]}_{-{\bf k}+{\bf p,\bf k}}&=&\mp \hat{t}_{(-{\bf k}+{\bf p})_z}(\hat {t}_{{\bf k}_x}+\hat{t}^{\dagger}_{-{\bf k}_x})\mp\hat{t}_{(-{\bf k}+{\bf p})_x}(\hat {t}_{{\bf k}_z}+\hat{t}^{\dagger}_{-{\bf k},z})\\&&+ i\hat{t}_{({-\bf k}+{\bf p})_z}(\hat {t}_{{\bf k}_y}+\hat{t}^{\dagger}_{-{\bf k}_y})+ i\hat{t}_{(-{\bf k}+{\bf p})_y}(\hat {t}_{{\bf k}_z}+\hat{t}^{\dagger}_{-{\bf k},z})\nonumber \\
\hat {T}^{[0]}_{-{\bf k}+{\bf p,\bf k}}&=&\sqrt{\frac{2}{3}}\big[-\hat{t}_{(-{\bf k}+{\bf p})_x}(\hat {t}_{{\bf k}_x}+\hat{t}^{\dagger}_{-{\bf k}_x})\\&&-\hat{t}_{(-{\bf k}+{\bf p})_y}(\hat {t}_{{\bf k}_y}+\hat{t}^{\dagger}_{-{\bf k}_y})+2 \hat{t}_{(-{\bf k}+{\bf p})_z}(\hat {t}_{{\bf k}_z}+\hat{t}^{\dagger}_{-{\bf k}_z}) \big]
\end{eqnarray*}
From Eq.~(\ref{pertubationEk}) one can notice that, within the second order perturbation theory, only the non-zero matrix element will contribute to the correction energy for a state $|n\rangle$ of three quasiparticle excitations. This states will be composed of a local quintet excitations (with an energy cost $= 2J^\prime+\lambda$) and two triplet excitations. After using the Bogolibov transformation of triplets in Eq.~(\ref{pertubationEk}), and computing the non-zero matrix elements in new two triplon quasiparticle state, $\hat{\gamma}^{\dagger}_{{\bf k_1}\alpha}\hat{\gamma}^{\dagger}_{{\bf k_2}\beta}|0\rangle$, such as $\langle{\bf k}_{1\alpha},{\bf k}_{2\beta} | (\bf {k-p})_\alpha,{\bf -k_\beta}\rangle \ne0$. The summation term in Eq.~(\ref{pertubationEk}) for each $m$ can be written as, \\$\sum_{n\ne0}\sum_{\bf p,k}\frac{|\langle n| \hat {T}^{[m]}_{-{\bf k}+{\bf p},{\bf k}}\epsilon_{\bf k}|0\rangle|^2}{E_0-E_n}$ 
\begin{eqnarray*}
=-2\sum_{\bf p,k}\Big[\sinh^2\theta_{\bf k}\epsilon^2_{-{\bf k}+{\bf p}}(\cosh2\theta_{-{\bf k}+{\bf p}}-\sinh2\theta_{-{\bf k}+{\bf p}})\nonumber\\
+\sinh^2\theta_{{-\bf k}+{\bf p}}\epsilon^2_{{\bf k}}(\cosh2\theta_{{\bf k}}-\sinh2\theta_{{\bf k}})\Big]\nonumber\\
/\big(2J^\prime+\lambda+E_{\bf -k}+E_{{\bf k}-{\bf p}}\big).
\end{eqnarray*}
The ground state energy per dimer including second order correction can be written:
\begin{eqnarray}
 e_{g} \big[\lambda,\bar{s}^{2}\big]&=&e_{0}+\frac{3}{2N_{d}}\sum_{\bf{k}} E_{\bf{k}}+\Delta E^S_{\bar{s}\hat{t} \hat{t}\hat{q}}
 \end{eqnarray}
Minimising $e_g$ with respect to $\bar s^2$ and $\lambda$, we have calculated the critical points for transition from dimerised phase to N\'eel  phase. We find the values of critical points as:  $J_c\approx$[0.153, 0.08, 0.05, 0.034, 0.025, 0.009] for S=1, 3/2, 2, 5/2, 3 and 5, respectively.  These values approximately consistent with  Darradi \emph{et al}~\cite{RDarradi2005} calculations and lie in between their variational mean-field like calculation and the coupled-cluster method, by which they were able to calculate the critical points only for $S\le3/2$. For comparison their values of $J_c$ through CCM: $J_c =0.156$ ($S=1$) and $J_c=0.091$ ($S=3/2$). Notably, the values of critical points from our perturabtive calculation with the inclusion of quintet states reduces approximately around 20-25$\%$, with respect to the critical values obtained from the  minimal triplon mean-field theory. But one finds it difficult to generate the phase boundaries continuously in the whole parameter space using the second order perturbation theory. We, therefore, find a better and easier way to include quintets into the triplon mean-field theory.} 
\subsection{\color{black} Extended triplon analysis with quintet mean-fields: Improved quantum phase diagram}\label{stq-mft}
For a neater presentation, let us first redefine the quintet operators in the real spherical-harmonics form.
\begin{eqnarray}
\hat{q}^\dag_{xy} &=& \frac{i}{\sqrt{2}}(\hat{q}^\dag_{\bar{2}} - \hat{q}^\dag_2) ,\quad\hat{q}^\dag_{x^2-y^2} = \frac{1}{\sqrt{2}}(\hat{q}^\dag_{\bar{2}} + \hat{q}^\dag_2)\nonumber \\ 
\hat{q}^\dag_{z^2} &=& \hat{q}^\dag_0,\quad\hat{q}^\dag_{xz} = \frac{1}{\sqrt{2}}(\hat{q}^\dag_{\bar{1}} - \hat{q}^\dag_1) ,\quad \hat{q}^\dag_{yz} = \frac{i}{\sqrt{2}}(\hat{q}^\dag_{\bar{1}} + \hat{q}^\dag_1)\nonumber
\end{eqnarray}
With an $\sbar$ describing the singlet, the $\Hhat_{\that\that\qhat}$ type terms being linear in $\qhat$'s will induce some static values for the quintet operators in the mean-field approximation. Hence, we introduce five real quintet mean-field variables, $\bar{q}_{z^2}$, $\bar{q}_{x^2-y^2}$, $\bar{q}_{xy}$, $\bar{q}_{yz}$ and $\bar{q}_{xz}$. Now the intra-dimer exchange interaction can be written as:
\begin{eqnarray}
 {\bf S}_1\cdot{\bf S}_2 && = -S(S+1) \bar{s}^2 + \{[-S(S+1) + 3]\times \nonumber \\ 
&&(\bar{q}^2_{z^2} + \bar{q}^2_{x^2-y^2} + \bar{q}^2_{xy}+ \bar{q}^2_{yz} + \bar{q}^2_{xz} )\}  \nonumber \\ 
&&+ [-S(S+1) + 1] \sum_\alpha \hat{t}^\dagger_\alpha\hat{t}^{ }_\alpha.
 \end{eqnarray}
 Moreover, the constraint can be written as: 
 \begin{equation}
\bar{s}^2 + (\bar{q}^2_{z^2} + \bar{q}^2_{x^2-y^2} + \bar{q}^2_{xy} + \bar{q}^2_{yz} + \bar{q}^2_{xz}) +  \sum_\alpha \hat{t}^\dag_\alpha\hat{t}^{ }_\alpha =1\nonumber \\ 
\end{equation}
As we are interested in the approach to classical behavior for increasing $S$, of all the terms in the bond-operator representation given Eqs.~(\ref{latticeS1}) and~(\ref{latticeS2}), we only keep those with the highest power of $S$. After approximating the singlet and quintet operators by their mean-fields, we get the following simplified representation of the spin operators.
\begin{equation}
\left[\begin{array}{c} S_{1x}\\ S_{1y}\\ S_{1z}\end{array}\right] \approx \left\{ \sbar \sqrt{\frac{\tt N_t}{2 {\tt N}_s}} \mathbb{I}^{ } + \frac{1}{2}\sqrt{\frac{{ \tt N_q}}{\tt{ \tt N_t}}} \Qbar \right\} 
\left[\begin{array}{c} \that^\dag_x+\that^{ }_x \\ \that^\dag_y+\that^{ }_y \\ \that^\dag_z+\that^{ }_z\end{array}\right] 
\end{equation}
\begin{equation}
\left[\begin{array}{c} S_{2x}\\ S_{2y}\\ S_{2z}\end{array}\right] \approx - \left\{ \sbar \sqrt{\frac{{\tt N}_t}{2{\tt N}_s}} \mathbb{I}^{ } + \frac{1}{2}\sqrt{\frac{{\tt N}_q}{{\tt N}_t}} \Qbar \right\} 
\left[\begin{array}{c} \that^\dag_x+\that^{ }_x \\ \that^\dag_y+\that^{ }_y \\ \that^\dag_z+\that^{ }_z\end{array}\right] 
\end{equation}
Here, $\mathbb{I}$ is the $3\times 3$ identity matrix, and 
\[ \Qbar = \left[\begin{array}{ccc} 
\frac{2}{\sqrt{3}} \qbar_{xx} & \qbar_{xy} & \qbar_{xz} \\
\qbar_{xy} & \frac{2}{\sqrt{3}} \qbar_{yy} & \qbar_{yz} \\
\qbar_{xz} & \qbar_{yz}  & \frac{2}{\sqrt{3}} \qbar_{zz}
\end{array}\right]. \]
 Moreover, $\qbar_{xx} + \qbar_{yy} + \qbar_{zz}=0$, where $\qbar_{zz} = \qbar_{z^2}$, $\qbar_{xx} = -(\qbar_{z^2} - \sqrt{3}\,\qbar_{x^2-y^2})/2$ and $\qbar_{yy} = -(\qbar_{z^2} + \sqrt{3}\,\qbar_{x^2-y^2})/2.$ 
 
 The quintet-corrected effective triplon dynamics of the Heisenberg model on staggered-dimer square lattice is now given by the following Hamiltonian.
\begin{eqnarray} \label{eq:hmfsttq_kspc}
\tilde{H}_{t}&= &N_d\Big[J^\prime-J^\prime S(S+1)-\frac{5}{2}\lambda+\bar{s}^2\big(\lambda-J^\prime\big) +\big(2J^\prime+\lambda\big)\nonumber\\ &&\times\big(\bar q^2_{x^2-y^2}+\bar q^2_{z^2}\big)\Big] +\frac{\lambda}{2}\sum_{\bf{k},\alpha} \left(\hat t^\dag_{\bf{k}\alpha} \hat t_{\bf{k} \alpha} + \hat t_{-\bf{k}\alpha}\hat t^\dag_{-\bf{k}\alpha}\right) \nonumber\\ &&
-\frac{1}{2}\sum_{\bf{k},\alpha,\beta} \epsilon_{\bf k} \Bigg\{\left(\that^\dag_{\k,\alpha}+\that^{ }_{-\k,\alpha}\right) \times \nonumber \\&& \left[ \sbar \sqrt{\frac{ {\tt N}_t}{2{\tt N}_s}}\mathbb{I}^{ } + \frac{1}{2}\sqrt{\frac{{\tt N}_q}{{\tt N}_t}} \Qbar \right]^2_{\alpha\beta} \left(\that^\dag_{-\k,\beta}+\that^{ }_{\k, \beta}\right)\Bigg\}
\end{eqnarray}
Here, with unit lattice spacing, $\epsilon_{\bf k}=2J\big[\cos (2k_{x})+2\cos(k_{x})\cos(k_{y})\big]-8J_f\cos(k_{x})\cos(k_{y})$. The triplon Hamiltonian, $\tilde{H}_t$, can be simplified by diagonalising the real symmetric matrix, $\bar{s} \sqrt{\frac{{\tt N}_t}{2{\tt N}_s}} \mathbb{ I}^{ } + \frac{1}{2}\sqrt{\frac{{\tt N}_q}{{\tt N}_t}} \bar{Q}$, which is purely dependent on $\bar{s}$ and $\bar{q}$'s. Under this diagonalisation, the three component triplon vector would also rotate canonically. In terms of these new triplons, the different values of $\alpha (=x,y,z)$ do not mix with each other, and one can  study $\tilde{H}_t$ in the same way as $\Hhat_t$ of Eq.~\ref{eq:hmf_kspc}. Note that, due to the tracelessness of $\Qbar$, its three eigenvalues would have the same general structure as its diagonal elements, $\qbar_{xx}$ etc. Hence, without any serious loss of generality, we consider a simpler case with $\bar{q}_{xy}=\bar{q}_{xz} =\bar{q}_{yz}=0$.

The ground state energy per dimer, $e_g$, of $\tilde{H}_t$
can be written as:
\begin{eqnarray}
e_g&=&e_0+\big(2J^\prime+\lambda\big)\big(\bar{q}^2_{z^2}+\bar{q}^2_{x^2-y^2}\big)+\frac{1}{2N_d}\sum_kE_{\k_{x}}\nonumber\\
&&+\frac{1}{2N_d}\sum_\k E_{\k_{y}}+\frac{1}{2N_d}\sum_\k E_{\k_{z}}, \label{eq:eg-stq}
\end{eqnarray}
where the three quasiparticle dispersions are gives as follows.
\begin{eqnarray*}
 E_{\k,{\alpha}}=\sqrt{\lambda^2-2\lambda\epsilon_\k\left(\sqrt{\frac{{ \tt N}_t}{2{\tt N}_s}} \bar s+\sqrt{\frac{ {\tt N}_q}{3{\tt N}_t}} \bar q_{\alpha\alpha}\right)^2} 
\end{eqnarray*}
Note that the triplon dispersions here will in general be non-degenerate. The presence of $\qbar$'s will pull one or more of these dispersions lower, which will eventually become gapless at some ${\bf Q}$ to cause magnetic order. We will find below that, indeed, this improved triplon dynamics exhibits stronger tendency towards ordering.
\begin{figure*}[ht]
\begin{center}
\includegraphics[height=2.25in]{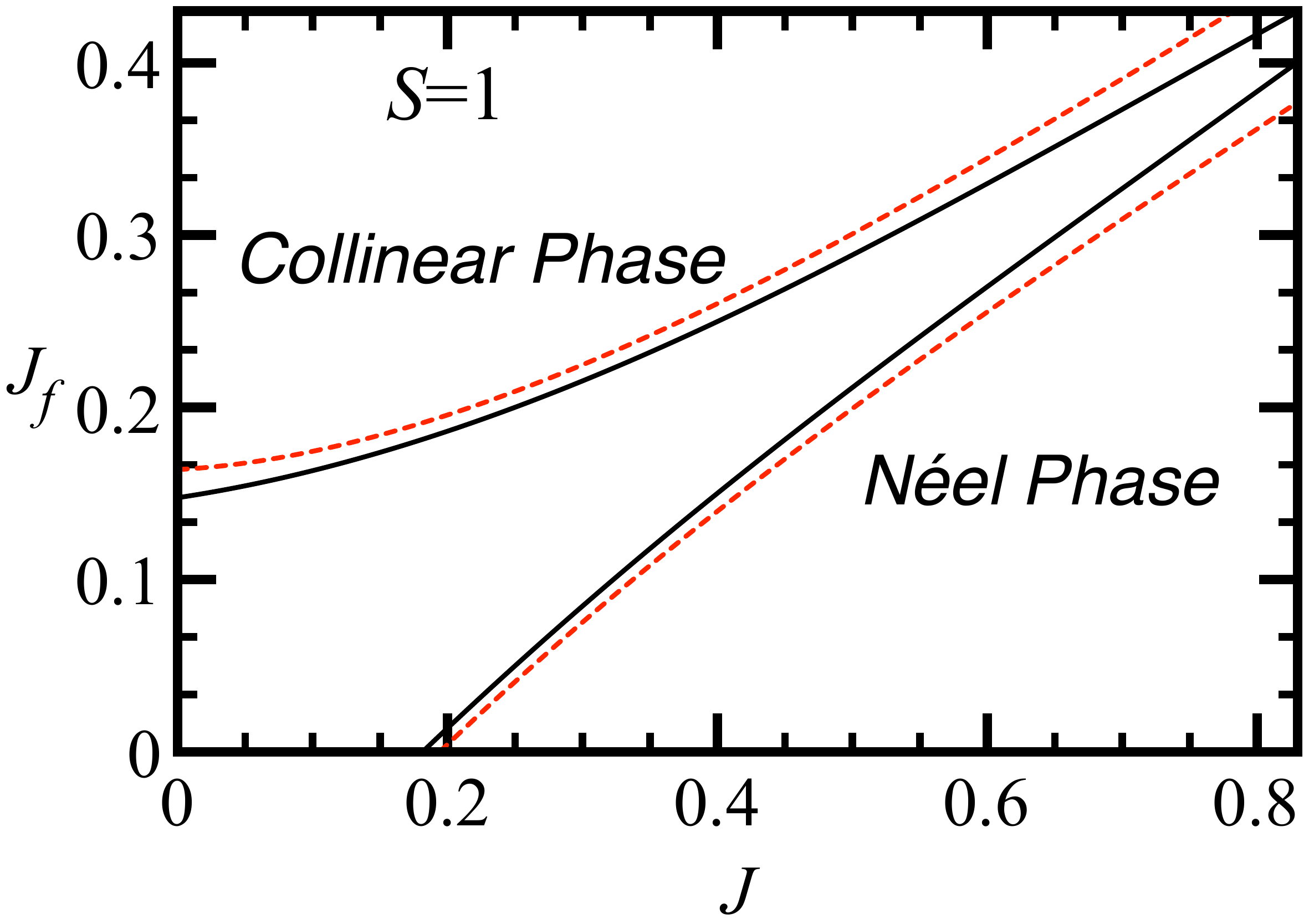}
\includegraphics[height=2.25in]{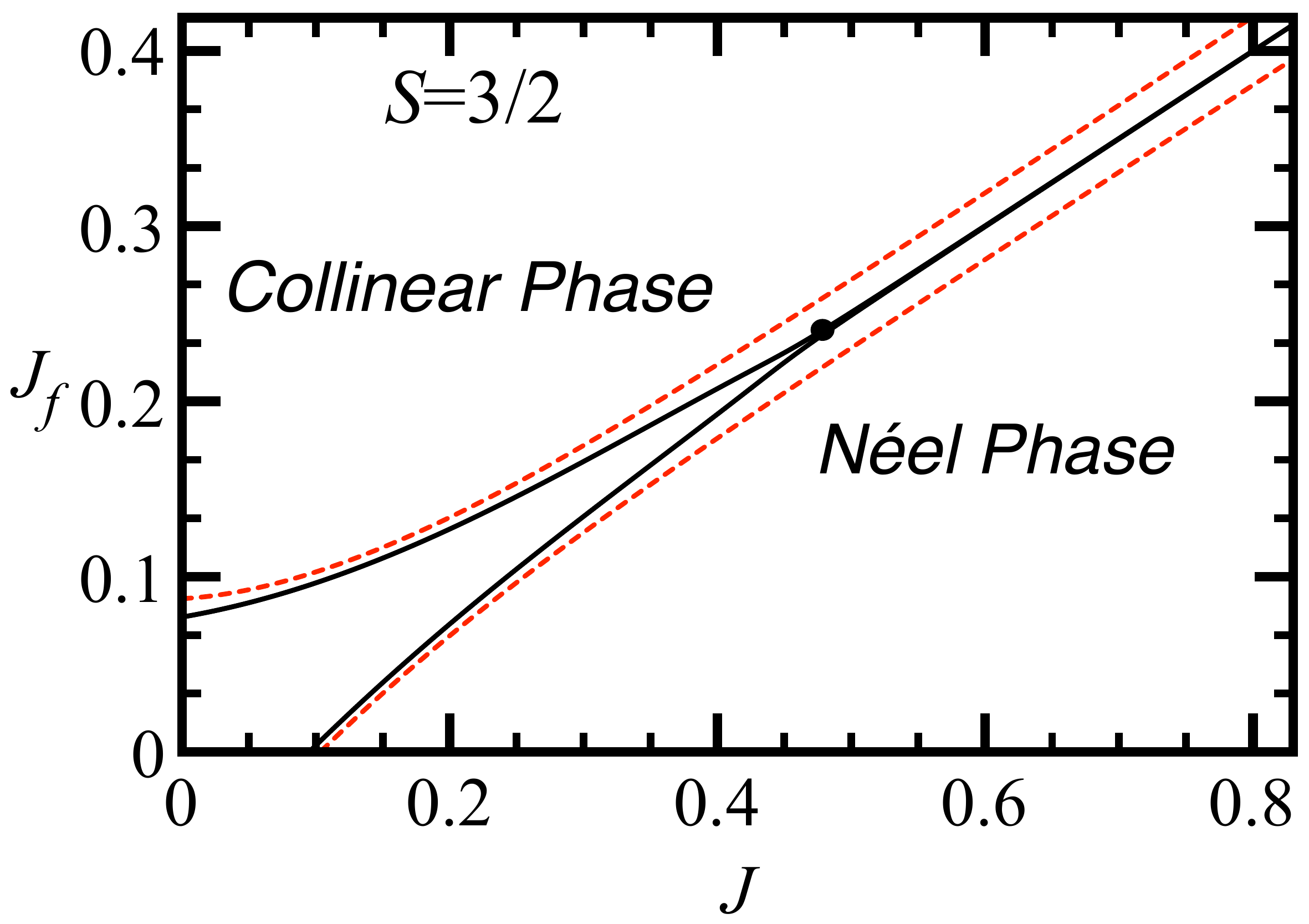}\\
\includegraphics[height=2.25in]{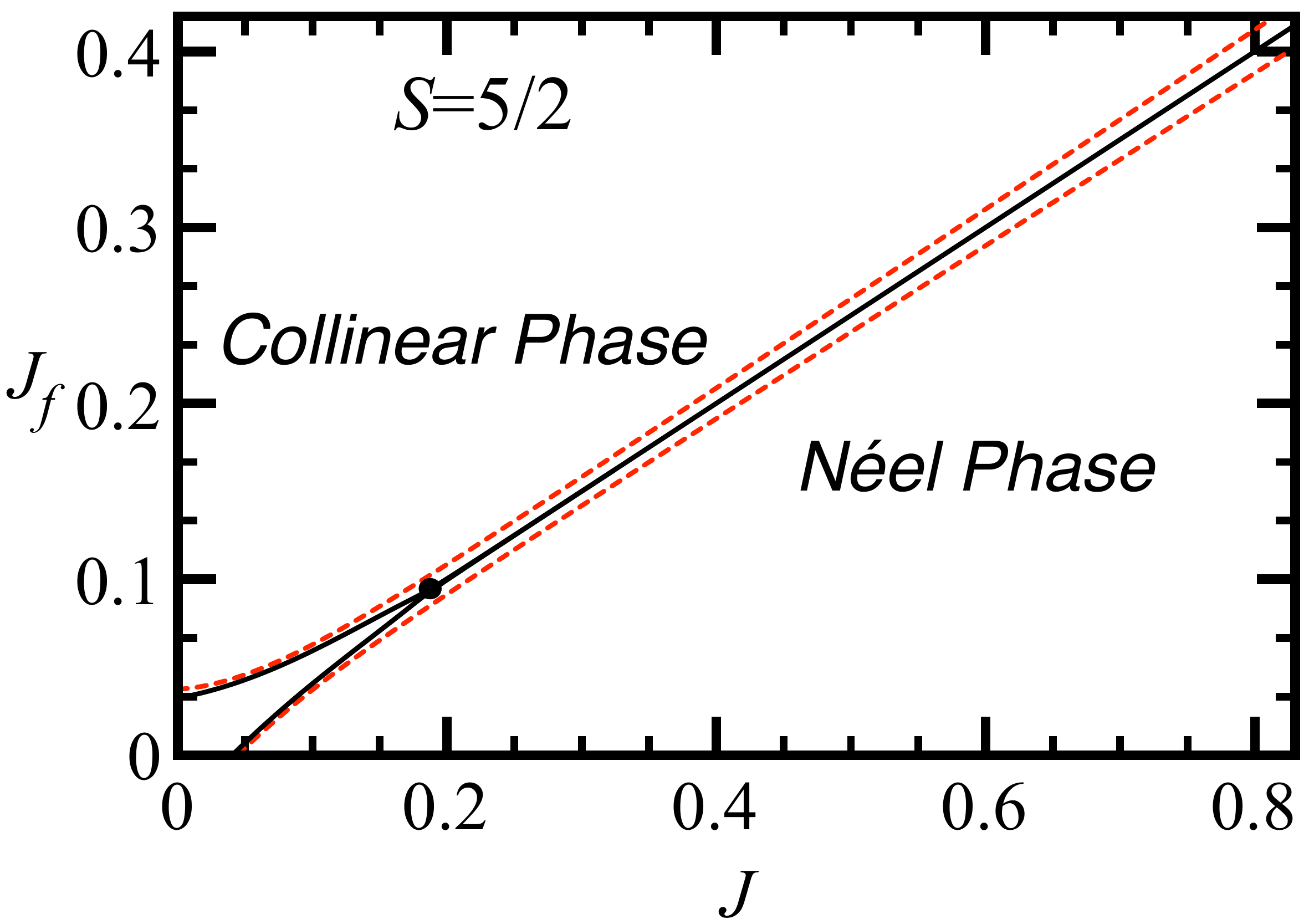}
\includegraphics[height=2.25in]{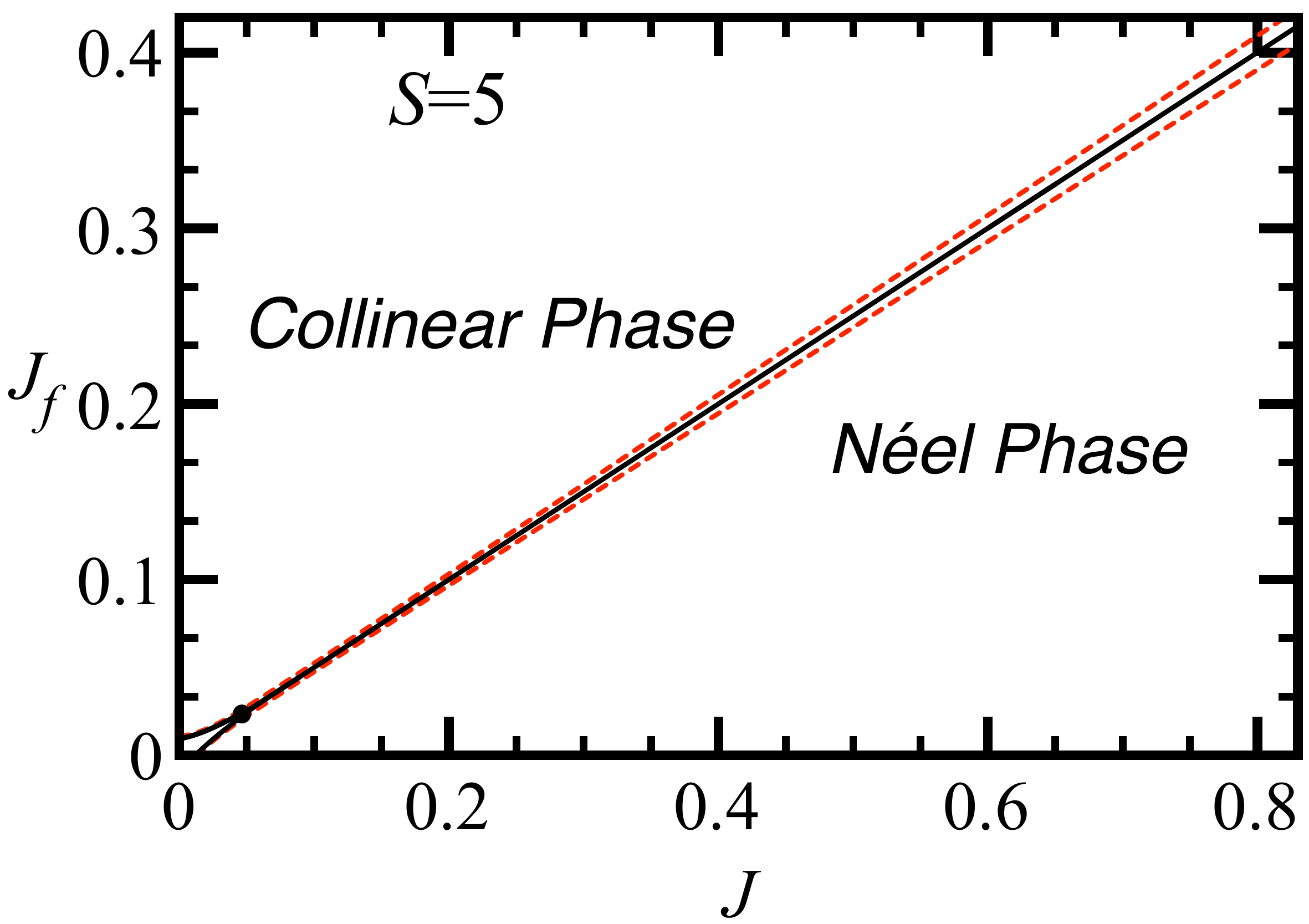}
\caption {The quantum phase diagrams obtained from the extended triplon analysis for different values of $S$. The black lines represents the phase boundaries obtained from the extended triplon analysis including quintets, and, the red lines are the phase boundaries obtained from the minimal triplon analysis shown in Fig.2. The boundaries of the N\'eel and collinear phases merge into the Majumdar-Ghosh line, $J_f=J/2$, at the tricritical point (filled black circle), $J_{tc}$.
}
\label{mergeline_stq}
\end{center}
\end{figure*}

Minimising the $e_g$ of Eq.~(\ref{eq:eg-stq}) with respect to $\lambda$, $\bar{s}$, $\bar q_{z^2}$ and $\bar q_{x^2-y^2}$ gives us the following mean-field equations.
\begin{eqnarray*}
\bar{s}^2&=&\frac{5}{2}-\bar{q}^2_{z^2}-\bar{q}^2_{x^2-y^2} \nonumber\\
&&-\frac{1}{2N_d}\sum_\k\frac{\lambda-\epsilon_\k\left(\sqrt{\frac{{\tt N}_t}{2{\tt N}_s}} \bar s+\sqrt{\frac{{\tt N}_q}{3{\tt N}_t}} \bar q_{xx}\right)^2}{E_{\k_x}} \nonumber\\
&&-\frac{1}{2N_d}\sum_\k\frac{\lambda-\epsilon_\k\left(\sqrt{\frac{{\tt N}_t}{2{\tt N}_s}} \bar s+\sqrt{\frac{{\tt N}_q}{3{\tt N}_t}} \bar q_{yy}\right)^2}{E_{\k_y}} \nonumber\\
&&-\frac{1}{2N_d}\sum_\k\frac{\lambda-\epsilon_\k\left(\sqrt{\frac{{\tt N}_t}{2{\tt N}_s}} \bar s+\sqrt{\frac{{\tt N}_q}{3{ \tt N}_t}} \bar q_{zz}\right)^2}{E_{\k_z}}
\end{eqnarray*}
\begin{eqnarray*}
\lambda&=&J^\prime+\frac{\lambda}{2N_d}\Big(\frac{\tt N_t}{2\tt N_s}+\frac{1} {\bar s}\sqrt{\frac{\tt N_q}{6\tt N_s}}\bar q_{xx}\Big)\sum_\k\frac{\epsilon_\k}{E_{\k_x}}  \nonumber\\
&&+\frac{\lambda}{2N_d}\left(\frac{\tt N_t}{2\tt N_s}+\frac{1} {\bar s}\sqrt{\frac{\tt N_q}{6\tt N_s}}\bar q_{yy}\right)\sum_\k\frac{\epsilon_\k}{E_{\k_y}}\nonumber \\&&+\frac{\lambda}{2N_d}\left(\frac{\tt N_t}{2\tt N_s}+\frac{1} {\bar s}\sqrt{\frac{\tt N_q}{6\tt N_s}}\bar q_{zz}\right)\sum_\k\frac{\epsilon_\k}{E_{\k_z}}
\end{eqnarray*}
\begin{eqnarray*}
\bar{q}_{zz}&=&-\frac{1}{4N_d}\Big(\frac{\lambda}{2J^\prime+\lambda}\Big) \left(\frac{\tt N_q}{3\tt N_t}\bar{q_{xx}}+\bar{s}\sqrt{ \frac{\tt N_q}{6\tt N_s}}\right) \sum_\k\frac{\epsilon_\k}{E_{\k_x}}  \nonumber\\
&&-\frac{1}{4N_d}\Big(\frac{\lambda}{2J^\prime+\lambda}\Big) \left(\frac{\tt N_q}{3\tt N_t}\bar{q_{yy}}+\bar{s}\sqrt{ \frac{\tt N_q}{6\tt N_s}}\right) \sum_\k\frac{\epsilon_\k}{E_{\k_y}}  \nonumber\\
&&+\frac{1}{2N_d}\Big(\frac{\lambda}{2J^\prime+\lambda}\Big) \left(\frac{\tt N_q}{3\tt N_t}\bar{q_{zz}}+\bar{s}\sqrt{ \frac{\tt N_q}{6\tt N_s}}\right) \sum_\k\frac{\epsilon_\k}{E_{\k_z}} 
\end{eqnarray*}
\begin{eqnarray*}
\bar{q}_{x^2-y^2}=\frac{1}{4N_d}\Big(\frac{\sqrt{3}\lambda}{2J^\prime+\lambda}\Big) \left(\frac{\tt N_q}{3\tt N_t}\bar{q_{xx}}+\bar{s}\sqrt{ \frac{\tt N_q}{6\tt N_s}}\right) \sum_\k\frac{\epsilon_\k}{E_{\k_x}}\nonumber\\
-\frac{1}{4N_d}\Big(\frac{\sqrt{3}\lambda}{2J^\prime+\lambda}\Big) \left(\frac{ \tt N_q}{3 \tt N_t}\bar{q_{yy}}+\bar{s}\sqrt{ \frac{\tt N_q}{6\tt N_s}}\right) \sum_\k\frac{\epsilon_\k}{E_{\k_y}}.
\end{eqnarray*} 
To generate the phase boundaries between the staggered-dimer and AFM phases, we solve the above self-consistent mean-field equations for $\lambda$, $\bar{s}^2$, $\bar{q}_{z^2}$ and $\bar{q}_{x^2-y^2}$, iteratively. We find the minimum ground state energy solution to have $\bar{q}_{zz}\ne0$ and $\bar{q}_{x^2-y^2}=0 $. This leads to the lowering of the dispersion, $E_{{\bf k}_z}$. 
 
\begin{figure}[htbp]
\centering
\includegraphics[height=2.in]{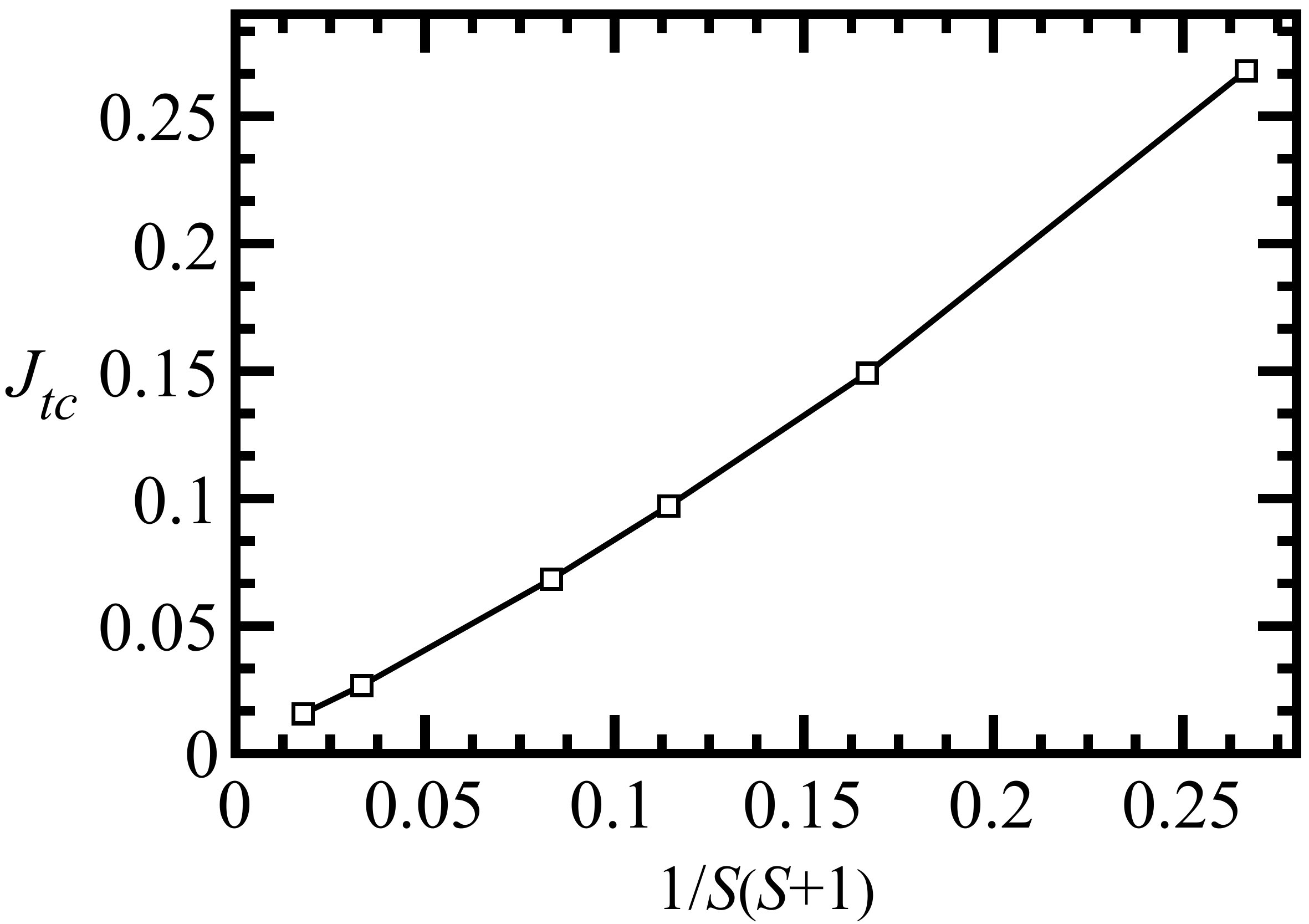}
\caption {The tricritical point, $J_{tc}$, versus $1/S(S+1)$.  The $J_{tc}$ is defined as the distance from the origin of the point of merger of the N\'eel and collinear phase boundaries in Fig.~\ref{mergeline_stq}.} 
\label{criticalmergeline}
\end{figure}

The quantum phase diagrams obtained from this approach for different values of $S$ are shown in Fig~\ref{mergeline_stq}. The phase boundaries are generated for the ordering wavevector $\Q$ =$(0,0)$ for the N\'eel phase, and $\Q$ =$(0,\pi)$ for the collinear phase, with $E_{{\bf Q}z}=0$, which fixes the chemical potential at $\lambda^{*}=2\epsilon_\Q \Big(\sqrt{\frac{N_t}{2N_s}} \bar s+\sqrt{\frac{N_q}{3N_t}} \bar q_{zz}\Big)^2$. Here, we find the staggered-dimer phase to shrink rapidly, as the spin quantum number $S$ increases.  For $S\ge 3/2$, the two critical phase boundaries of the staggered-dimer phase are found to merge at the tricritical point, $J_{tc}$, beyond which the N\'eel and collinear phases are directly separated from each other by the well-known Majumdar-Ghosh line, $J_f/J=1/2$~\cite{Pchandra1988,Danu2016}. As plotted in Fig.~\ref{criticalmergeline}, the $J_{tc}$ tends to zero as $S$ increases. Notably, for numerical comparison at $J_f=0$, our improved critical $J_c$ for $S=1$ is $0.18141$, which nearly same as the exact diagonalization estimate of $J_c=0.1818$ for S=1, and for $S=3/2$ it is 0.0951 in our calculation which lies in between the values of $J_c= 0.099,0.0917$ estimated from ED and CCM, respectively (for $S=3/2$) in Ref.~\cite{RDarradi2005}. It is satisfying to see that our extended triplon mean-field theory is able to get the consistent physical behaviour while approaching the classical limit from a quantum state. We close this section by briefly noting that this approach for dimerized quantum antiferromagnets can be extended to include the septets, octets etc. by treating the even total-spin bond-operators in mean-field, while keeping the odd total-spin bond-operators as such. It would to lead an effective bilinear probelm in the odd spin bond-operators, which can be studied in the same manner as done here.
\section{Staggered-dimer model including single-ion anisotropy} 
\label{Model with single-ion anisotropy}
 Let us first understand a single antiferromagnetic dimer with axial spin anisotropy.
\begin{eqnarray}\label{dimerD}
\Hhat_{\mbox{\tiny dimer}} &=&J^{\prime} {\bf S}_{1}.{\bf S}_{2}-D({\bf S}^{2}_{1z}+{\bf S}^{2}_{2z}).
\end{eqnarray}
A non-zero $D$ causes splitting and mixing between the total-spin states of a dimer. For instance, now the singlet mixes with the triplet as well as the quintet states. This we can see by constructing ${\bf S}^2_{1z}$ and ${\bf S}^2_{2z}$ in the subspace of singlet, triplet and quintet states. Thus, in the bond-operator representation, 
 \begin{eqnarray}
 {\bf S}^2_{(1,2)z}& \approx&\sqrt{\frac{{\tt N}_q}{6{\tt N}_{s}}}\left(\shat^{\dagger}\qhat_{0}+\qhat^{\dagger}_{0}\shat\right) +\left({\frac{{\tt N}_{t}}{2{\tt N}_{s}}}\right)\shat^{\dagger}\shat \nonumber \\&&+\left({\frac{{\tt N}_t}{2{\tt N}_s}}+{\frac{{ \tt N}_q}{3{\tt N}_t}}\right)\that^{\dagger}_{0}\that_{0} +{\frac{1}{4}}\left({\frac{{\tt N}_q}{{\tt N}_t}}+1\right)\left(\that^{\dagger}_1\that_1+\that^{\dagger}_{\bar{1}}\that_{\bar {1}}\right)   \nonumber \\&&\pm\sqrt{\frac{{\tt N}_q}{{2\tt N}_t}}\left(\that^{\dagger}_1\qhat_1+\qhat^{\dagger}_1\that_1-\that^{\dagger}_{\bar{ 1} }\qhat_{\bar {1}}-\qhat^{\dagger}_{\bar{1}}\that_{\bar{1}}\right)
  \nonumber \\
 &&+{\frac{1}{21}}\left(11S^{2}+11S-15\right)\qhat^{\dagger}_{0}\qhat_{0}  \nonumber \\&&+{\frac{1}{14}}\left(6S^{2} +6S-5\right)\left(\qhat^{\dagger}_{1}\qhat_{1}+\qhat^{\dagger}_{\bar {1}}\qhat_{\bar{ 1}}\right)
  \nonumber \\
 &&+{\frac{1}{7}}\left(S^{2}+S+5\right)\left(\qhat^{\dagger}_{2}\qhat_{2}+\qhat^{\dagger}_{\bar{ 2}}\qhat_{\bar {2}}\right). 
 \label{eq:D-dimer}
 \end{eqnarray}
Notice the mixing between $\shat$ and $\qhat_{0}$ in Eq.~(\ref{eq:D-dimer}), which in addition to the triplons, will most directly affect the dimer singlet state. Hence, we  now write $\Hhat_{\mbox{\tiny dimer}}$ of Eq.~(\ref{dimerD}) in terms of $\shat$, $\that_\alpha$ and $\qhat_0$ only, and ignore all the other bond-operators for simplicity. In this approximation, the $\Hhat_{\mbox{\tiny dimer}}$ takes the following form.
\begin{eqnarray}
&\Hhat_{\mbox{\tiny dimer}}\approx-J'S(S+1)\Big(\shat^{\dagger}\shat+\that^{\dagger}_{\alpha}\that_{\alpha}+\that^{\dagger}_z\that_z+\qhat^{\dagger}_0\qhat_0\Big) \nonumber \\
 & +JÕ\Big(\that^{\dagger}_{\alpha}\that_{\alpha}+\that^{\dagger}_z\that_z+3 \qhat^{\dagger}_0\qhat_0\Big)-D\left({\frac{{\tt N}_t}{{\tt N}_s}}\right)\shat^{\dagger}\shat \nonumber \\
 &-D \sqrt{\frac{2{\tt N}_q}{3{\tt N}_s}}\Big(\shat^{\dagger}\qhat_0+\qhat^{\dagger}_0\shat \Big)- 2D \Big({\frac{{\tt N}_{t}}{2 {\tt N}_s}}+{\frac{{\tt N}_q}{3{\tt N}_t}}\Big)\that^{\dagger}_z\that_z \nonumber \\
 &-\frac{D}{2}\Big({\frac{{\tt N}_q}{{\tt N}_t}}+1\Big)\that^{\dagger}_{\alpha}\that_{\alpha}-{\frac{2D}{21}}\left(11S^{2}+11S-15\right)\qhat^{\dagger}_0\qhat_0 \label {bndhml}
\end{eqnarray}
where $\alpha=x,y$. The mixing of $\shat$ with $\qhat_0$ in Eq.~(\ref{bndhml}) can be diagonalized by unitary rotation of $\shat$ and $\qhat_0$ to new operators $\tilde{s}$ and $\tilde{q}_0$ such that $\shat=\tilde{s} \cos{\theta}-\tilde{q}_{0} \sin{\theta}$ and $\qhat_0=\tilde{s} \sin{\theta}+\tilde{q}_0 \cos{\theta}$. 
Here, \[ \tan 2 \theta={2D\sqrt {\frac{2{\tt N_q}}{3{\tt N_s}}}} / {\big[3J^{\prime}+D{\frac{{\tt N_t}}{{\tt N_s}}}\\-\frac{2D}{21}(11S^2+11S-15)\big]}.\] The $\Hhat_{\mbox{\tiny dimer}}$ in terms of the bond operators $\tilde{s}$, $\that_\alpha$ and $\tilde{q}_0$ has the following diagonal form. 
\begin{eqnarray}
\Hhat_{\mbox{\tiny dimer}} &\approx& \frac{1}{2}\Big(\tilde{s}^{\dagger}\tilde{s}+\tilde{q}^{\dagger}_{0}\tilde{q}_{0}\Big)\Big[-2J^{\prime}S(S+1)+3J^{\prime}-D\frac{\tt N_t}{\tt N_s}\nonumber\\
&&-\frac{2D}{21}\big(11S^2+11S-15\big)\Big]\nonumber\\
&&+\frac{1}{2}\left(-\tilde{s}^{\dagger}\tilde{s}+\tilde{q}^{\dagger}_{0}\tilde{q}_{0}\right) \times \Bigg(\frac{2D}{\sin 2\theta}\sqrt {\frac{2{\tt N_q}}{3{\tt N_s}}}\Bigg)\nonumber \\ 
&&-2D \Big({\frac{{\tt N}_{t}}{2 {\tt N}_s}}+{\frac{{\tt N}_q}{3{\tt N}_t}}\Big)\that^{\dagger}_z\that_z-\frac{D}{2}\Big({\frac{{\tt N}_q}{{\tt N}_t}}+1\Big)\that^{\dagger}_{\alpha}\that_{\alpha}
\label{dgnlzdbndhml}
\end{eqnarray} 
\subsection{\label{modified-rep} Modified bond-operator representation}
\begin{figure}[htbp]
\centering
 \includegraphics[width=2.75in]{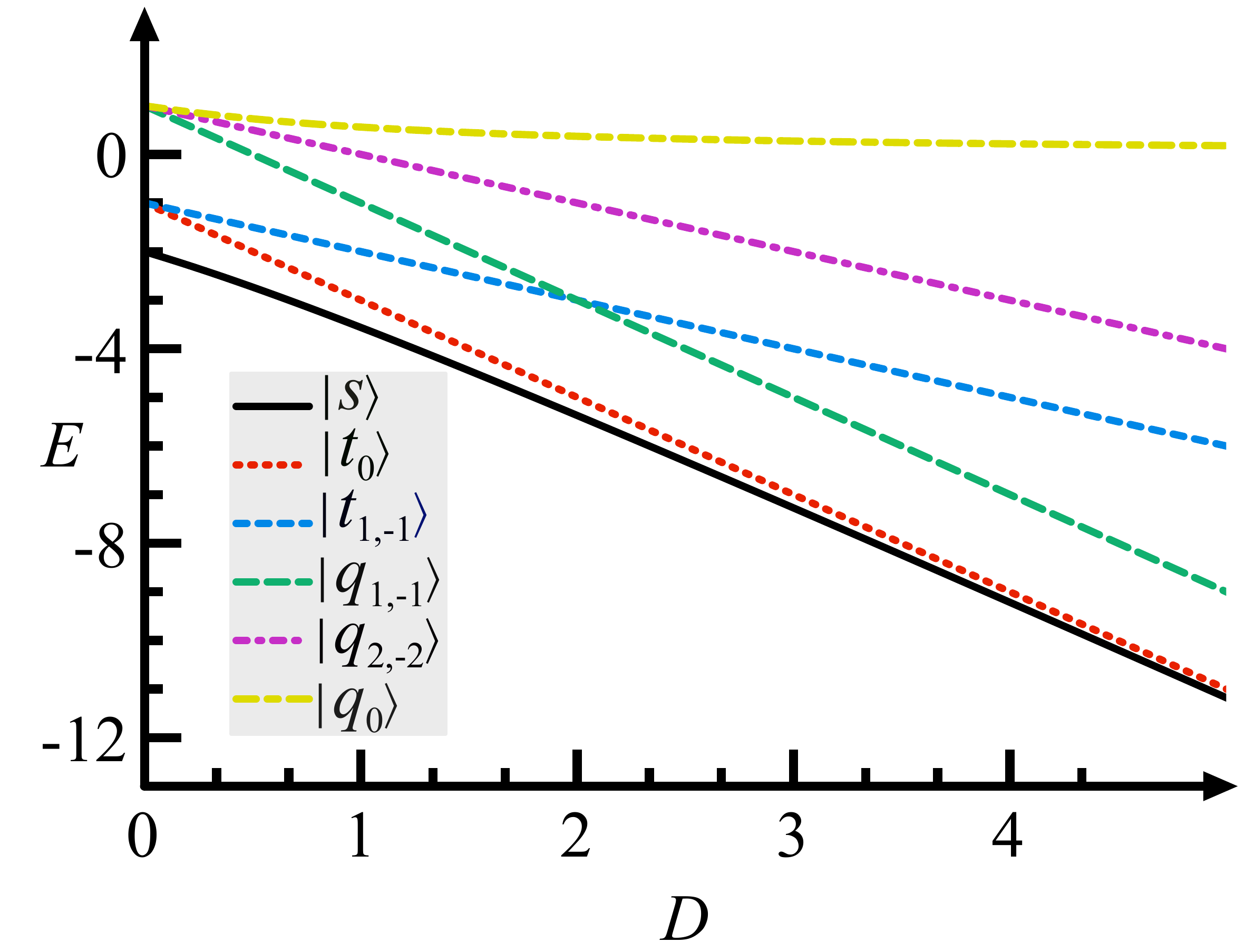}
 \caption {Energy levels splittings of dimer bond with the single ion anisotropy for $S$$=$1.}
 \label{fig:energy_s1}
\end{figure}

 The eigenvalues of the diagonalized dimer Hamiltonian of Eq.~(\ref{dgnlzdbndhml}) as a function of $D$ are plotted in Fig.~\ref{fig:energy_s1}. As the quintet state, $|\tilde q_0\rangle$, is much higher in energy, for a low-energy effective description, we can ignore it. Moreover, in the mean-field approximation $\tilde s$  is replaced by $\bar s$. Therefore, we get
 \begin{eqnarray}
&& \Hhat_{\mbox{\tiny{dimer}}} \approx \frac{1}{2}\bar{s}^{2}+\Big[-2J^{\prime}S(S+1)+3J^{\prime}-D\frac{\tt N_t}{\tt N_s}\nonumber\\
&&-\frac{2D}{21}(11S^2+11S-15) -\frac
{2D}{\sin 2\theta}\sqrt {\frac{2{\tt N_q}}{3{\tt N_s}}} \Big]\nonumber\\
&&+\Big[-J^{\prime}S(S+1)+J^{\prime}+\frac{D}{2}\Big({\frac{{\tt N}_q}{{\tt N}_t}}+1\Big)\Big]\hat{t}^{\dagger}_{\alpha}\hat{t_{\alpha}}\nonumber\\
&&+\Big[-J^{\prime}S(S+1)+J'+2D\Big({\frac{{\tt N}_t}{2{\tt N}_s}}+\frac{{\tt N}_q}{{\tt N}_t}\Big)\Big]\that^{\dagger}_z\that_z.
\label{dgnlzdbndhml_fnl}
\end{eqnarray} 
The bond-operator representation for ${\bf S}_{1\alpha}$ and ${\bf S}_{2\alpha}$ is also required to be reconstructed for this case. From Ref.~\cite{Kumar2010}, we know ${\bf S}_{1\alpha}$ and ${\bf S}_{2\alpha}$ in the restricted space of $\{|\shat \rangle$, $|\that_{\alpha}\rangle$, $|\that_z\rangle$, $|\qhat_0\rangle\}$. 
After taking care of the mixing between $\shat$ and $\qhat_{0}$, and then using the mean-field approximation for $\tilde s$ operator and ignoring the $\tilde q_0$ excitations, we get the following expressions for ${\bf S}_{1\alpha}$ and ${\bf S}_{2\alpha}$. 
\begin{eqnarray}
{\bf S}_{1\alpha} &=&-{\bf S}_{2\alpha}  \nonumber \\
&\approx&\bar{s} \left(\sqrt{\frac{ {\tt N}_t}{2{\tt N}_s}}\cos{\theta}-\frac{1}{2}\sqrt{\frac{ {\tt N}_q}{3{\tt N}_t}}\sin{\theta}\right)\big(\that^{\dagger}_{\alpha}+\that_{\alpha}\big) \label{eq:stq_fnl1}\\
{\bf S}_{1z} &=&-{\bf S}_{2z}  \nonumber \\
&\approx& \bar{s}  \left(\sqrt{\frac{ {\tt N}_t}{2{\tt N}_s}}\cos{\theta}+\sqrt{\frac{ {\tt N}_q}{3{\tt N}_t}}\sin{\theta}\right)\big(\that^{\dagger}_z+\that_z\big) 
\label{eq:stq_fnl2}
\end{eqnarray}
Here, $\alpha=x,y$. Using this representation, now we can do the standard triplon analysis of the full staggered-dimer model of Eq.~(\ref{stg_mdl}) including the spin anisotropy, $D$. 
\subsection{Triplon analysis and magnetic order for $D\ne0$ }
 \label{Triplon mean-field calculation with single ion anisotropy effect:}
The triplon Hamiltonian for the $\Hhat$ of Eq.~(\ref{stg_mdl}), including single-ion anisotropy has the following form.
  \begin{eqnarray}
 H^D_{t}&=&e^D_{0}N_{d}+\frac{1}{2}\sum_{\bf{k},\alpha} \Bigg\{(\lambda_{\alpha}-\bar{s}^{2} \phi_\alpha \xi_{\bf{k}}) (\that^{\dagger}_{\bf{k}\alpha} \that_{\bf{k}\alpha}+h.c) \nonumber \\&& -\frac{1}{2}\bar{s}^{2} \phi_\alpha \xi_{\bf{k}} \big(\that^{\dagger}_{\bf{k}\alpha}\that^{\dagger}_{-\bf{k}\alpha}+h.c\big)\Bigg\} \nonumber \\
 && +\frac{1}{2}\sum_{\bf{k}} \Bigg\{(\lambda_z-\bar{s}^{2} \phi_z \xi_{\bf{k}})(\that^{\dagger}_{\bf{k} z} \hat{t}_{\bf{k} z}+h.c) \nonumber \\&&-\frac{1}{2}\overline{s}^{2} \phi_z \xi_{\bf{k}} \big(\that^{\dagger}_{\bf{k} z}\that^{\dagger}_{-\bf{k} z}+h.c\big)\Bigg\}.
\label{HmnfD}
\end{eqnarray}
Here, $\xi_{\bf k}=2J\cos (2k_{x}a)+ 4J\cos(k_{x}a) \cos(k_{y}a)\\-8J_{f}\cos(k_{x}a)\cos(k_{y}a)\),  and the 
parameters $\phi_\alpha$ and $\phi_z$ as functions of  $S$, $J^\prime$ and $D$ are defined as, $\phi_{\alpha=x,y}=(\sqrt{\frac{{\tt N_t}}{2{\tt N_s}}}\cos\theta\\-\frac{1}{2}\frac{1}{\sqrt{3}}\sqrt{\frac{{\tt N_q}}{{\tt N_t}}}\sin\theta)^2$ and $\phi_z=\big(\sqrt{\frac{{\tt N_t}}{2{\tt N_s}}}\cos\theta+\frac{1}{\sqrt{3}}\sqrt{\frac{{\tt N_q}}{{\tt N_t}}}\sin\theta\big)^2$.
Similarly, the spin anisotropy induced split effective chemical potentials are: $\lambda_{\alpha=x,y}=\lambda-\frac{D}{2}(\frac{{\tt N}_q}{{\tt N}_t}+1)$, and, $\lambda_{z}=\lambda-2D(\frac{{\tt N}_t}{2{\tt N}_s}+\frac{{\tt N}_q}{3{\tt N}_t})$, and
\begin{eqnarray}
e^D_0&=&-J^\prime S(S+1)+J^\prime-\frac{5}{2}\lambda+\frac{D}{2}\left(\frac{{\tt N}_q}{{\tt N}_t}+1\right)\nonumber\\
&&+D\left(\frac{{\tt N}_t}{2{\tt N}_s}+\frac{{\tt N}_q}{3{\tt N}_t}\right)+\frac{1}{2}\bar{s}^{2}\Big[JÕ+2\lambda-D\frac{{\tt N}_t}{{\tt N}_s}\nonumber\\&&
-\frac{2D}{21}(11S^2+11S-15)-\frac{2D} {\sin 2\theta}\sqrt {\frac{2{\tt N_q}}{3{\tt N_s}}}\Big]. \nonumber
\end{eqnarray}
After Bogoliubov diagonalization, we get 
 \begin{eqnarray}
\Hhat^D_{t}& =&e^D_0 N_d +\sum_{{\bf k},\alpha} E_{{\bf k \alpha}}\left(\gammahat^{\dagger}_{{\bf k}\alpha}\gammahat_{{\bf k}\alpha}+\frac{1}{2}\right)\nonumber\\
&&+\sum_{{\bf k}} E_{{\bf k} z}\left(\hat{\gamma}^{\dagger}_{{\bf k} z}\gammahat_{{\bf k} z}+\frac{1}{2}\right).
\end{eqnarray}
Here, $E_{{\bf k \alpha}}$$=$$\sqrt{\lambda_{\alpha} \left(\lambda_{\alpha}-2\bar{s}^{2} \phi_\alpha \xi_{{\bf k}} \right)}$ for $\alpha=x,y$ and $E_{{\bf k } z}=\sqrt{\lambda_{z} \left(\lambda_{z}-2\bar{s}^{2} \phi_z \xi_{{\bf k}} \right)}$ are the quasiparticle dispersions. The ground state energy per dimer is
 \begin{eqnarray}
 e^D_{g}\big[\lambda,\bar{s}^2\big]&=&e^D_{0}+\frac{1}{2N_d}\sum_{{\bf k}, \alpha} E_{{\bf k \alpha}}+\frac{1}{2N_d}\sum_{{\bf k}} E_{{\bf k} z}.
\end{eqnarray}
Minimizing, $e^D_{g}, $ with respect to ${\lambda}$ and $\bar{s}^{2}$ gives the following mean-field equations.
\begin{figure*}[ht]
\begin{center}
\includegraphics[height=2.2in]{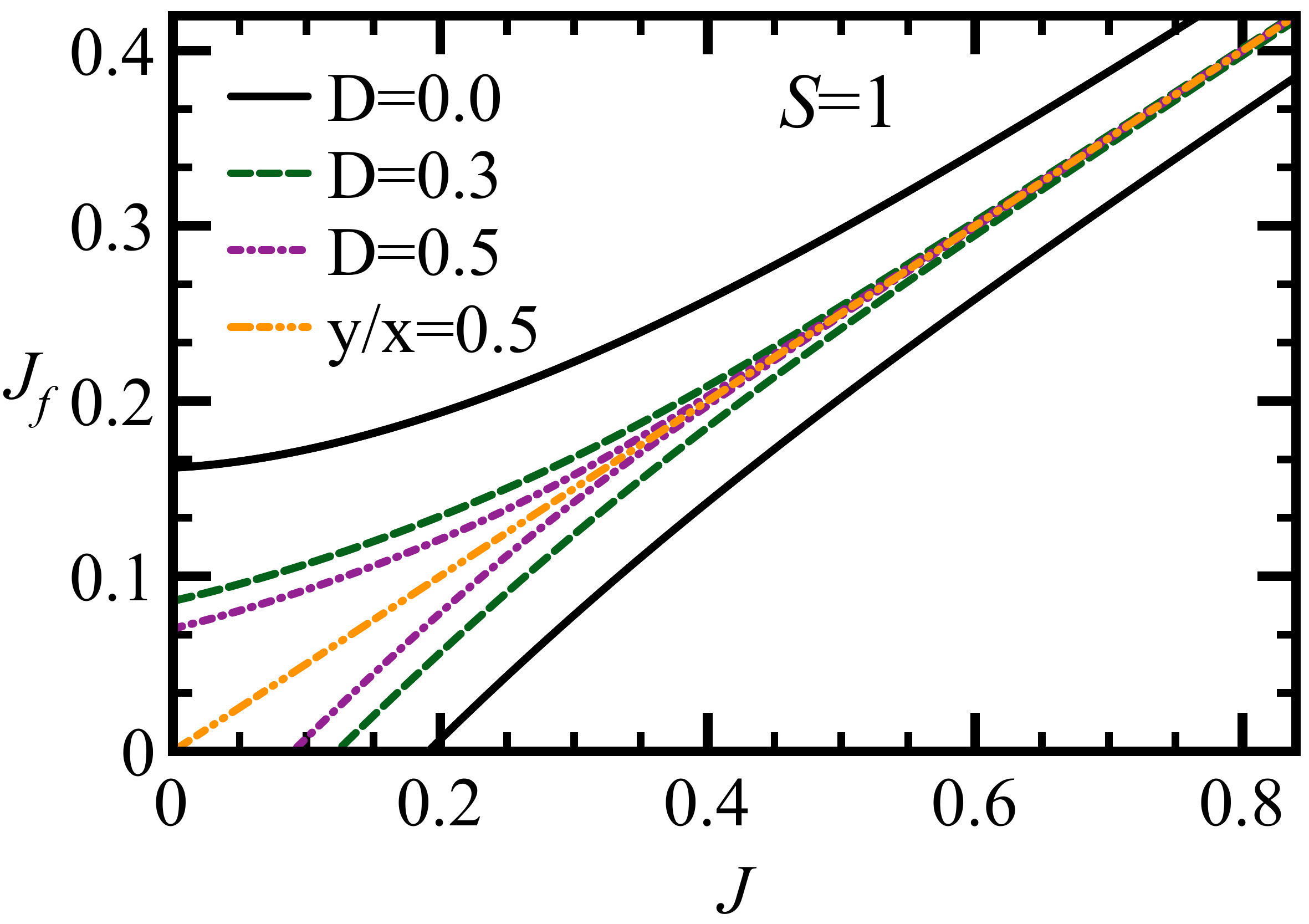}
\includegraphics[height=2.2in]{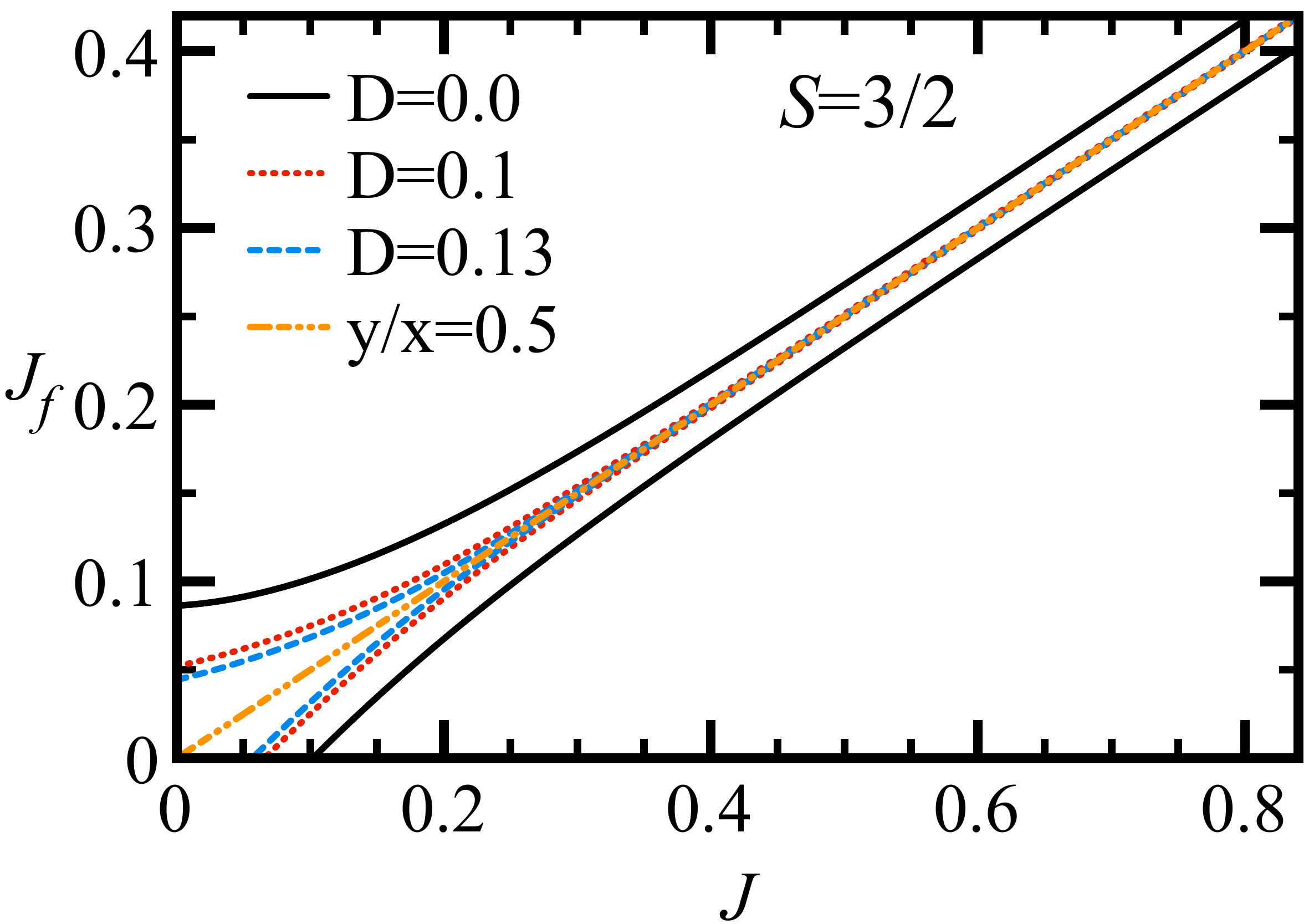}\\
\includegraphics[height=2.2in]{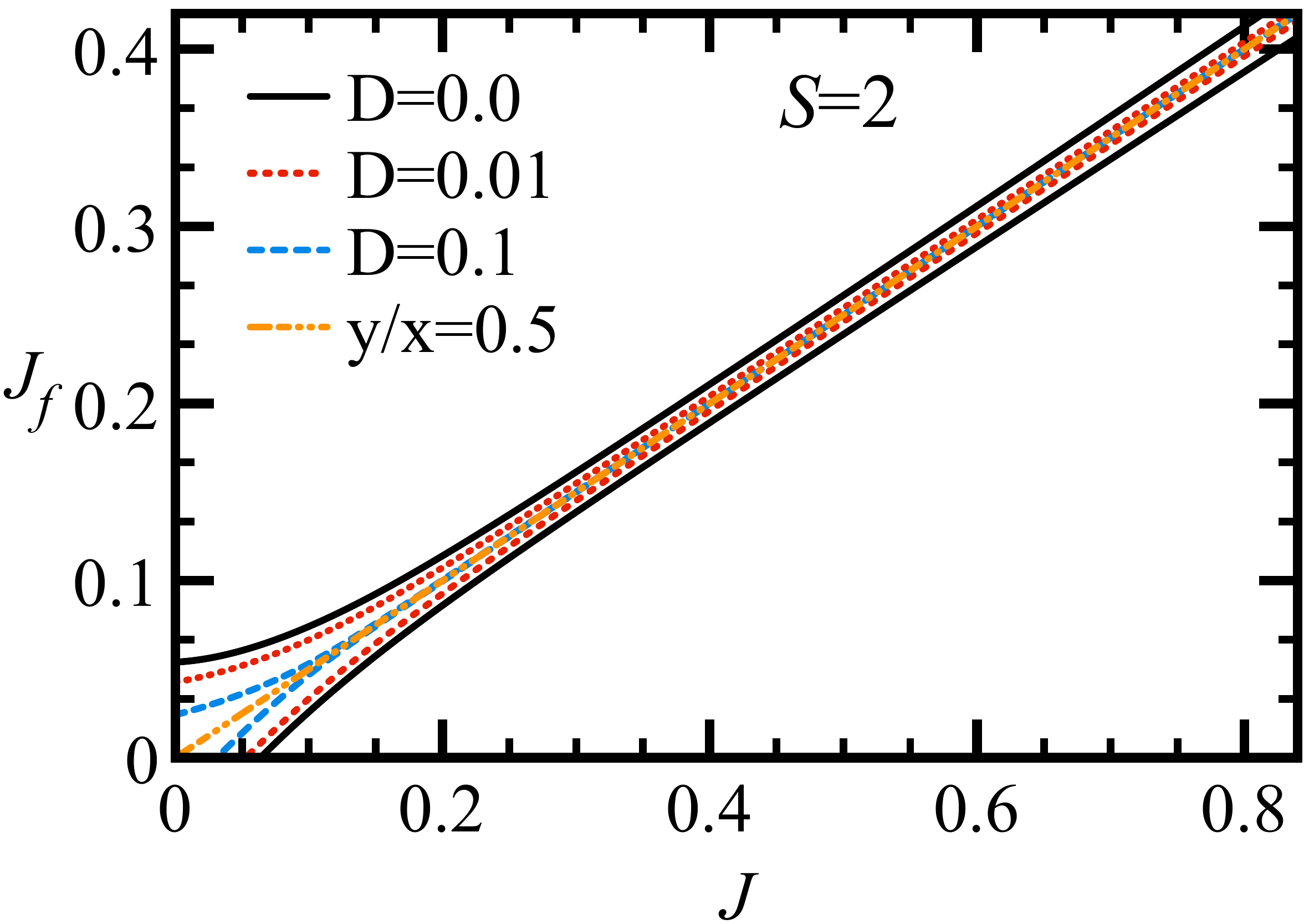}
\includegraphics[height=2.2in]{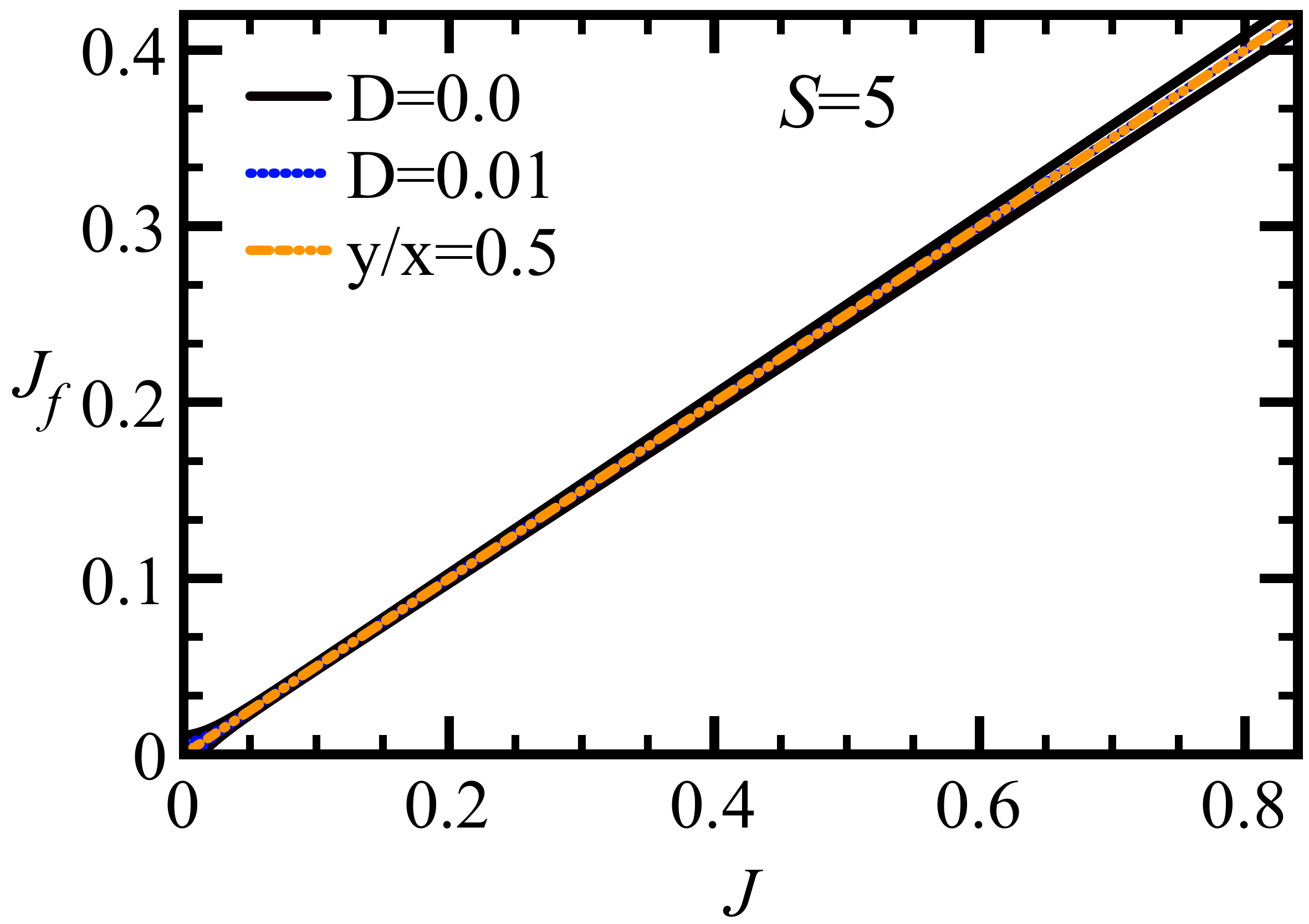}
\caption {The quantum phase diagram with the single ion anisotropy for $S=1$ (top-left), for $S=3/2$ (top-right), for $S=2$ (bottom-left), and for $S=5$ (bottom-right). The boundaries of the N\'eel and collinear phases merge into a single line at $J_{f}$$=$$0.5J$ for strong enough single-ion anisotropy ($D$).
 }
\label{mergeline}
\end{center}
\end{figure*}
 \begin{eqnarray}
\bar{s}^{2}&=&\frac{5}{2}-{\frac{1}{2N_d}}\Bigg\{\sum_{{\bf k}, \alpha}\frac{ \lambda_{\alpha}-\bar{s}^{2} \phi_\alpha \xi_{{\bf k}}}{ E_{{\bf k \alpha}}}-\sum_{{\bf k}}\frac{\lambda_z-\bar{s}^{2} \phi_z \xi_{{\bf k}}}{E_{{\bf k} z}}\Bigg\}
\label{Dsbar}\\
\lambda&=&-\frac{J^{\prime}}{2}+D\frac{{\tt N_t}}{{2\tt N_s}}+\frac{D}{21}(11S^2+11S-15)+\frac{D} {\sin 2\theta}\sqrt {\frac{2{\tt N_q}}{3{\tt N_s}}}  \nonumber \\
&& +{\frac{1}{2N_d}}\sum_{{\bf k}, \alpha}\frac{\lambda_{\alpha}\bar{s}^2 \phi_\alpha \xi_{{\bf k}}}{E_{{\bf k \alpha}}} +{\frac{1}{2N_d}}\sum_{{\bf k}}\frac{\lambda_z \bar{s}^2 \phi_z \xi_{{\bf k}}}{E_{{\bf k} z}}
\label {Dmu}
\end{eqnarray}
 
To generate the boundaries of the gapped staggered-dimer phase, we track the quasiparticle gap by solving Eqs.~(\ref{Dsbar})-(\ref{Dmu}) self-consistently for $\lambda$ and $\sbar^{2}$. Below we discuss the quantum phase diagram thus generated. 

\subsection{Quantum phase diagram for $D\ne 0$}
\label{Results and discussion:}
The closing of triplon gap at $\Q$ =$(0,0)$ in the N\'eel phase, and at $\Q$ =$(0,\pi)$ in the collinear phase occurs through $E_{{\bf Q}z}=0$, which fixes $\lambda_z$ at $\lambda^{*}_z=2\bar{s}^{2} \phi_z \xi_{\Q}$. This is similar to the dispersions of the Heisenberg case in Sec.~\ref{stq-mft}. For positive $D$, the $E_{{\bf k}\alpha}$ always remains gapped for $\alpha=x,y$.
  
\begin{figure}[htbp]
 \centering
 \includegraphics[width=2.75in]{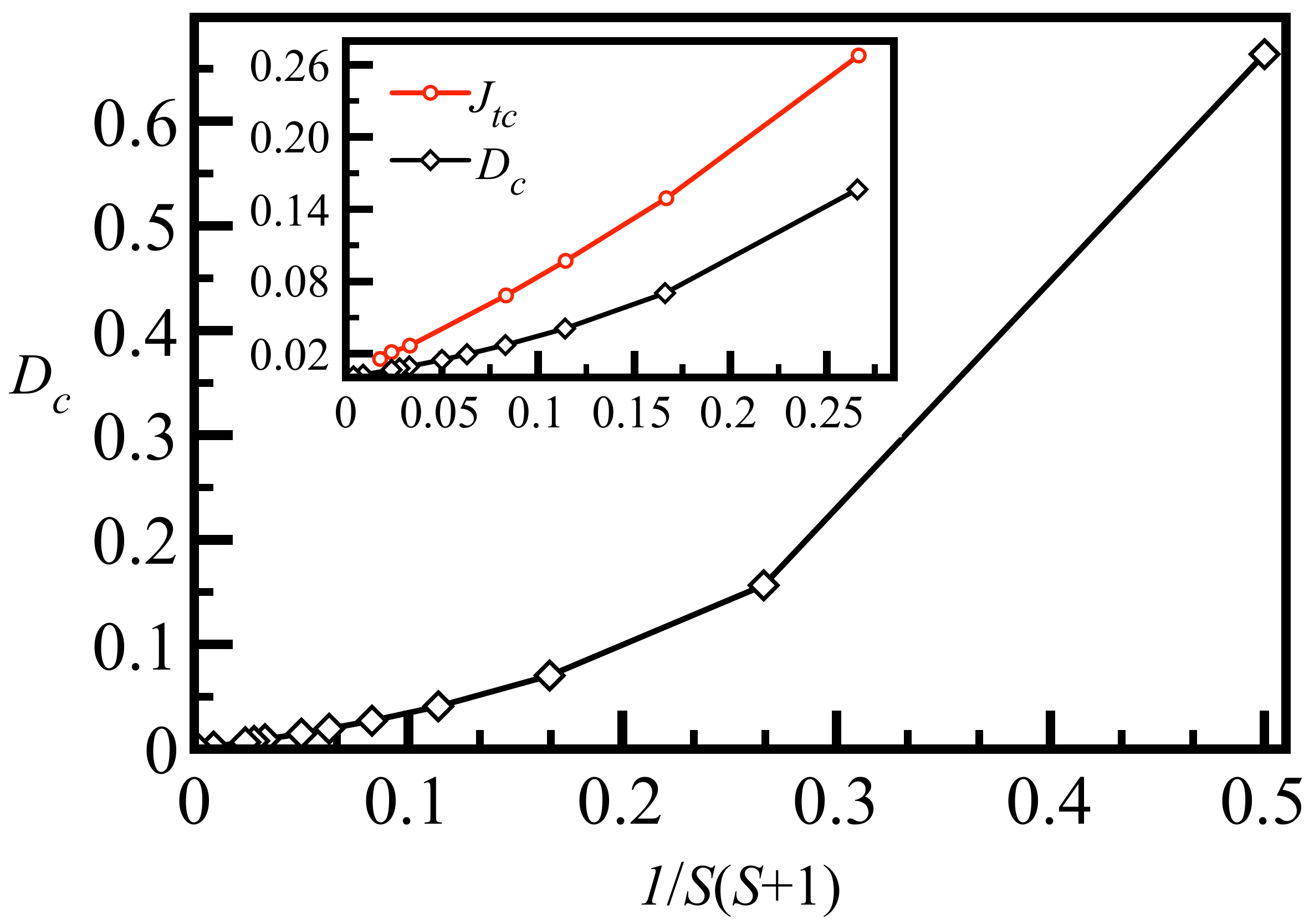}
 \caption {Critical values of $D_c$ as a function of spin quantum number $S$ from $S=1$ upto $S=15$. The  $J_{tc}$ and $D_c$  as function of $1/S((S+1)$ for merging two phase boundaries shown in subfigure. Both $J_{tc}$ and $D_c$ tends to zero as we approaches classical limit. } 
  \label{fig:d_vs_s(s+1)}
\end{figure}

  Recall that, for $D=0$, in the absence of quintet corrections we always found the dime phase, even for $S\rightarrow \infty$ (classical limit). But now, for $D>0$, we find that, above some critical value, $D_{c}$ at $J_{f}$$=$$0.5J$, the N\'eel and collinear phase boundaries merge into a single line for any value of $S$,  and the staggered-dimer phase completely vanishes. In Fig.~\ref{mergeline}, we show this for $S=1$, $S=3/2$, $S=2$ and $S=5$. 
  Thus, a sufficiently strong $D$ will always make the dimer phase disappear for any $S\ge1$.
We estimate the critical values of $D$ for first few values of $S$, which are $D_{c}\approx0.6633$ for $S=1$, $D_{c}\approx0.1563$ for $S=3/2$, $D_{c}\approx0.0720$ for $S=2$, and $D_{c}\approx0.0094$ for $S=5$.  We note that for large $S$, the critical $D_{c}$ becomes negligibly small, and the ground state is always a classical. In fact, $D_{c} \sim 1/S(S+1)$ for large $S$, as estimated from the data for upto $S=15$ in Fig.~\ref{fig:d_vs_s(s+1)}. This calculation clearly shows that, in the limit $S\rightarrow\infty$, an infinitesimal amount of single-ion anisotropy would suffice to completely suppress the quantum mechanical dimer phase, and to fully establish classical order.
\section {Conclusion:}
  \label{Summary:}
Here, we have studied the quantum phase transitions in a frustrated staggered-dimer model for arbitrary spin $S$ on square lattice by doing triplon mean-field theory, without and with single-ion anisotropy. First, we performed the minimal bond-operator mean-field theory in the reduced Hilbert space of singlet and triplets, without single-ion anisotropy. Through this, in the purely Heisenberg case, we found two ordered phases, viz. N\'eel and collinear states, sandwiching the spin-gapped staggered-dimer phase. The critical couplings for these transitions are found to scale precisely as $1/S(S+1)$. Puzzlingly, here we also found that this quantum (staggered-dimer) phase, helped by strong frustration, persists even in the limit $S\rightarrow\infty$ which is known to be a fully classical ordered state. This contradiction leads us to revise the singlet-triplet mean-field calculation by including the quintet states. To include quintet states first we performed a second order perturbation theory and present the improve critical points for transition from dimer to N\'eel phase. Next, we discuss the quintet participation in the Hamiltonian through a minimal singlet-triplet-quintet mean field theory. We find that this extended mean-field analysis is sufficient to consistently establish the classical AFM order for larger $S$. In fact through this, we find the N\'eel and collinear phases merge into a single line, $J_f=J/2 $, at the tricritical point $J_{tc} \sim 1/S(S+1)$.
 In our case, $J_f=J/2$ is the level-crossing line between two classical AFM phases.  Hence, we call the merger point, $J_{tc}$, as a tricritical point because it is a point where  the phase boundaries of t two continuous phase transitions meet with a line of  first order transition. Here, it is dimer to N\'eel, dimer to collinear, and, N\'eel to collinear phase transition. Thus, in general, infinitesimally small $J$ and $J_f$ would be sufficient to establish classical order in the limit of very large $S$.
In the presence of single-ion anisotropy, we first reworked the spin-$S$ bond-operator representation to include the effects  of singlet-quintet mixing on a dimer. Then, we used it to carry out the standard triplon analysis to work out the quantum phase diagram. Interestingly, for any $S \ge 1$, we always find a critical value of the anisotropy, $D_{c}$, for which the dimer phase between the N\'eel and collinear phases vanishes.  We also find that $D_{c} \sim 1/S(S+1)$ for large $S$. Thus, an infinitesimally small $D$ would be sufficient to kill the quantum mechanical dimer phase and establish classical order in the limit of very large $S$. Through this model calculation, we have presented an interesting perspective on the change of behaviour from quantum to classical in frustrated spin systems with increasing $S$. It would be nice to check this in other models, and if possible, using cold atoms in optical lattices~\cite{Alexey2011,Sarang2011,Simon2011,Briton2012}. The methods that we have developed here would, in general, be useful to dimerized quantum antiferromagnets of larger spins.
\section*{Acknowledgments} B. D. thanks R. Ganesh for useful discussions. B. K. acknowledges financial support under the UPE-II and DST-PURSE programs of JNU, New Delhi.

\end{document}